\documentclass[aps,prd, superscriptaddress,12pt]{revtex4-1}
\usepackage[colorlinks=true,allcolors=blue!92!black]{hyperref}
\usepackage{amssymb}
\usepackage{amsmath}
\usepackage{subfigure}
\usepackage{amsfonts}
\usepackage{xcolor}
\usepackage{epsfig}
\usepackage{color}
\usepackage{bbm}
\usepackage{bm}
\usepackage{tabularx}
\usepackage{multirow}
\usepackage{mathtools}
\usepackage{xfrac}
\usepackage{amsfonts}
\usepackage{latexsym}
\usepackage{amsmath,amssymb}
\usepackage{mathrsfs}
\usepackage{color,xcolor}
\usepackage{bm}
\usepackage{slashed}
\usepackage{url}
\usepackage{empheq}
\usepackage{textcomp}
\usepackage{dcolumn}
\usepackage{ulem}
\usepackage{cancel}
\usepackage{braket}
\usepackage[title]{appendix}
\usepackage{lipsum}
\usepackage{ulem}
\usepackage{cancel}
\usepackage{color}
\usepackage{cprotect}
\usepackage{orcidlink}
\usepackage{dblfnote}
\usepackage{cancel}
\usepackage[most]{tcolorbox}

\numberwithin{equation}{section} 
\def \l{\left}
\def \r{\right}
\def\beq{\begin{equation}}
	\def\eeq{\end{equation}}
\def\bea{\begin{eqnarray}}
	\def\eea{\end{eqnarray}}

\newcommand{\q}{\mathbf{q }}
\newcommand{\p}{\mathbf{ p}}
\newcommand{\x}{\mathbf{ x}}
\def\k{\textbf{k}}

\newcommand{\mean}[1]{\langle #1 \rangle}

\newcommand{\abs}[1]{\left|#1\right|}

\newcommand{\tr}[1]{\mathrm{Tr}#1}

\makeatletter

\def\inte{\textrm{int}}

\newcommand{\K}{\mathbf{ K}}

\def\k{\textbf{k}}

\makeatother

\begin{document}

	\title{ \Large  
		Exploring gravitational impulse via quantum Boltzmann equation
	}

	\author{M. Sharifian\,\orcidlink{0000-0001-9801-239X}
	}
	\email[]{mohammad.sharifian@uestc.edu.cn}
	\affiliation{Institute of Fundamental and Frontier Sciences, University of Electronic Sciences and Technology of China, Chengdu 611731, China}
	\affiliation{Key Laboratory of Quantum Physics and Photonic Quantum Information, Ministry of Education, University of Electronic Science and Technology of China, Chengdu 611731 , China}
	
	\author{M. Zarei\,\orcidlink{0000-0001-7744-2817}
	}
	\email[]{m.zarei@iut.ac.ir}
	\affiliation{Department of Physics, Isfahan University of Technology, 84156-83111 Isfahan, Iran}
	\affiliation{ICRANet-Isfahan, Isfahan University of Technology, 84156-83111 Isfahan, Iran}

	\author{N. Bartolo\,\orcidlink{0000-0001-8584-6037}
	}
	\email[]{nicola.bartolo@pd.infn.it}
	\affiliation{Dipartimento di Fisica e Astronomia “Galileo Galilei” Universita` di Padova, I-35131 Padova, Italy}
	\affiliation{INFN, Sezione di Padova, I-35131 Padova, Italy}
	\affiliation{INAF - Osservatorio Astronomico di Padova, I-35122 Padova, Italy}

	\author{S. Matarrese\,\orcidlink{0000-0002-2573-1243}
	}
	\email[]{sabino.matarrese@pd.infn.it}
	\affiliation{Dipartimento di Fisica e Astronomia “Galileo Galilei” Universita` di Padova, I-35131 Padova, Italy}
	\affiliation{INFN, Sezione di Padova, I-35131 Padova, Italy}
	\affiliation{INAF - Osservatorio Astronomico di Padova, I-35122 Padova, Italy}
	\affiliation{Gran Sasso Science Institute, I-67100 L'Aquila, Italy}

	
	\begin{abstract}
		\baselineskip=6mm
		
		We investigate the gravitational impulse by using the generalized formulation of the quantum Boltzmann equation (QBE), wherein the initial states are taken as wave packets rather than plane waves. The QBE operates within an open quantum system framework. Using this approach, we can analyze the two-body gravitational scattering by considering one body as part of the environment and the other as the system. Through this procedure, we obtain the system's impulse up to the second order for two massive objects. Similarly, we apply the forward scattering of the QBE to the gravitational impulse of a photon due to a massive object, during which we consider the photon as the system and the massive object as the environment. In this methodology, we use the QBE to compute the time evolution of the momentum operator associated with the system. While the forward scattering term vanishes when considering point particles, we demonstrate its persistence when using wave packets to describe both massive particles and the photon field.  The results reported here are entirely consistent with the previous approaches.

	\end{abstract}
	
	\maketitle
	

	\setcounter{page}{1}
	\newcounter{bean}

	\setcounter{tocdepth}{2}
	
	\newpage
	\tableofcontents

	\section{Introduction}
	
	As of today, it is widely accepted that classical general relativity can be interpreted as the $\hbar \rightarrow 0$ limit of a quantum mechanical path integral, where the action includes the Einstein-Hilbert term at a minimum \cite{Feynman:1963ax, DeWitt:1967yk,DeWitt:1967ub,DeWitt:1967uc}. In this framework, a gravitational interaction can be explained in terms of the exchange and interaction of spin-2 gravitons with matter and with each other.
	
	According to effective field theory, a quantum field theoretic description of gravity at large distances can be well-defined by using a derivative expansion order by order \cite{Donoghue:1993eb}. As a result, quantum mechanics indicates that higher-derivative terms should be added to classical general relativity. Additionally, what is typically viewed as a quantum mechanical loop expansion contains components of entirely classical nature, which can be of arbitrarily high order \cite{Iwasaki:1971vb, Holstein:2004dn}. In this case, it appears that factors of $\hbar$ are being canceled out, leading to the surprising conclusion that classical general relativity can be defined perturbatively by extending loops. In such a case, $\hbar$ serves as a dimensionful regulator, meaning it has units of measurement that influence the scale of the physical quantities, but it does not apply directly to classical physics. Instead, $\hbar$ regulates intermediate stages of the calculation or transition between quantum and classical descriptions, where quantum effects are relevant but classical physics is still being approximated~\cite{Iwasaki:1971vb, Holstein:2004dn}.
	
	Researchers have explored numerous approaches to conservative and radiative processes in 
	the gravitational two-body problem.
	These approaches encompass direct solutions of the General Relativity (GR) equations, either perturbative \cite{Kovacs:1978eu, Buonanno:1998gg, Damour:2000we, Damour:2001bu, Blanchet:2013haa, Barack:2018yvs,Abrahao2024the,Shojaei2023constraining}
	or numerical, e.g.~\cite{Pretorius:2005gq,Bishop:2016lgv}; effective field theory \cite{Goldberger:2004jt,Goldberger:2009qd,Foffa:2011ub,Foffa:2012rn,Levi:2015msa,Foffa:2016rgu,Levi:2018nxp,Endlich:2016jgc}; and
	use of the eikonal approximation \cite{Torgerson:1966zz,Abarbanel:1969ek,Levy:1969cr,Saotome:2012vy,Ciafaloni:2018uwe,KoemansCollado:2019ggb,Bern:2020gjj,Parra-Martinez:2020dzs,DiVecchia:2021bdo,Adamo:2021rfq,Porto2025nonlinear}. Progress on scattering amplitudes have also revitalized the examination of gravitational interactions within the framework of fully  relativistic formalism, see for example references~\cite{Damour:2017zjx,Damour:2019lcq,Kalin:2020fhe,Kalin:2020lmz,Damour:2020tta,Dlapa:2021npj,Bini:2021gat}. An approach to the effective one-body description of gravitationally interacting two-body systems has been introduced by Damour~\cite{Damour:2016gwp}. Furthermore, the scattering of two massive particles represented by scalar fields with wavefunctions has been described by Kosower \textit{et al.}~\cite{Kosower:2018adc}. The on-shell formalism in~\cite{Kosower:2018adc} offers a robust framework for bridging the gap between quantum and classical physics through scattering amplitudes. It finds practical applications in calculating gravitational waveform and impulse during astrophysical events. 
	
	Here, we take a different route and show
	that the QBE method which is built on the open quantum system approach can be used to derive classical results of gravity. The QBE is a generalization of the classical Boltzmann equation which describes the evolution of the distribution function of a gas of classical particles \cite{Chapman1991the,Cercignani2012boltzmann}. In contrast, the QBE describes the behavior of a gas of quantum particles in the presence of an external field. The QBE can be formulated as an open quantum system in which the behavior of quantum systems is studied in the presence of an environment.  
	The QBE is derived by assuming that the system is weakly coupled to the environment and also the time scale for the evolution of the system is much larger than the time scale of the environment \cite{kosowsky1995cosmic,Bartolo:2018igk,Bartolo:2019eac,Hoseinpour:2020hic,Sharifian2024open,Bavarsad:2009hm}.
	Recent developments in theoretical cosmology have led to the application of an open quantum system approach to the early universe, resulting in the development of a new (non-Markovian) quantum Boltzmann equation. This equation can be used to study a wide variety of processes including investigating the generation of circular polarization in cosmic microwave background (CMB)~\cite{kosowsky1995cosmic,Bavarsad:2009hm,Bartolo:2018igk,Bartolo:2019eac,Hoseinpour:2020hic}, damping of gravitational waves in an ultra-relativistic neutrino medium~\cite{Zarei:2021dpb}, the decoherence induced by squeezed gravitational waves~\cite{Sharifian2024open,Colas2024decoherence,Yoshimura2024decoherence}, and probing axion like particles~\cite{Sharifian2024induced,Hajebrahimi2023axion}.
	
	In this work, we use QBE to investigate the gravitational impulse by a massive object on another massive object and also on a light beam. In this framework, we consider one massive object as the environment and the other one as the system. In the case of the photon, the photon beam is considered as the system.  By considering point particles the first term of QBE (which corresponds to the forward scattering term) vanishes. However, we show that this term can survive when the fields are considered as wavefunctions.  For two massive objects, we also calculate the second order contribution corresponding to the collision term of the QBE.   After that, we perform a trace over the environmental degrees of freedom, along with determining the expectation value of the system's creation and annihilation operators. We see that the impulse results are in complete agreement with Damour~\cite{Damour:2016gwp} and Kosower \textit{et al.}~\cite{Kosower:2018adc}. We will show that the limit under which we derive the impulse using QBE is equivalent to the Goldilocks zone previously discussed in the literature.
	
	This manuscript has been organized as follows. In Section~\ref{sec:review} the description of two formalisms about the gravitational impulse with a quantum mechanical approach is presented. After that, in Section~\ref{sec:massive} we suggest the usage of QBE in the calculation of gravitational impulse. We see that the consideration of the wave packet prevents the momentum operator evolution to be vanished (at the lowest order). Using this QBE approach in Section~\ref{sec:light} we discuss the gravitational deflection of light beams induced by massive objects. Finally, the conclusion is provided in Section~\ref{sec:conc}.

	
	\section{Review of gravitational impulse}\label{sec:review}

	In the gravitational scattering of two particles, a significant observable is the net change in the initial momentum along a generic direction. The gravitational impulse is the difference between the incoming and outgoing momenta. Various methods have been suggested for calculating the gravitational impulse of particles from their scattering amplitude~\cite{Portilla1979momentum,Ledvinka2008relativistic,Kosower:2018adc,Damour:2020tta}. In the following, we review two of these approaches.

	\subsection{ KMOC formalism}
	
	The formalism, presented by Kosower, Maybee, and O'Connell (KMOC) \cite{Kosower:2018adc}, serves as a suggested bridge connecting quantum observables to classical quantities using the framework of scattering amplitudes. This formalism also enables the computation of gravitational waveforms emitted during the dynamic interactions of massive objects, e.g. binary black holes or neutron star mergers \cite{Porto:2016pyg,Kosower:2022yvp}. Furthermore, the on-shell formalism facilitates the calculation of gravitational impulse, which is the momentum exchange between merging objects via the gravitational wave. The impulse on a specific particle, denoted as $\Delta p^\mu=p'^\mu-p^\mu$, can be expressed as the difference between its initial, $p^\mu$, and final momentum, $p'^\mu$. They consider two wave packets initially separated by a transverse impact parameter $b^\mu$. In this scattering process, they quantify gravitational impulse of particle 1, $\Delta p^\mu$,  using a corresponding momentum operator $\mathbb{P}$, as expressed by the equation
	\begin{equation}
		\Delta p^\mu = \langle \text{out} | \mathbb{P} | \text{out} \rangle - \langle \text{in} | \mathbb{P} | \text{in} \rangle~.
	\end{equation}
	To gain access to the ``out" state, the $S$ matrix, serving as the evolution operator, is employed. This matrix encapsulates the time evolution of the system from the distant past to the far future. The ``out" state is connected to the ``in" state via the relationship $\ket{\text{out}} = S\ket{\text{in}}$. An alternative way to express the change in the observable is using the scattering amplitude that is related to the $S$ matrix by $\mathcal{A} = -i(S-1)$. In the classical limit, they obtain the leading-order impulse as
	\begin{eqnarray}\label{eq:KMOC}
		\Delta p^{\mu,(0)} =i \frac{\kappa^2}{4}\frac{\hbar }{(2\pi)^2}\int d^4 \bar{K} \delta\left(\bar{K} \cdot p\right) \delta\left(\bar{K} \cdot q\right) e^{-i b \cdot \bar{q}} \bar{K}^\mu \bar{\mathcal{A}}^{(0)}\left(p, q \rightarrow p+\hbar \bar{K}, q-\hbar \bar{K}\right) .~~~
	\end{eqnarray}
	where $\kappa=\sqrt{32\pi G}$ with $G$ as the gravitational constant, $q^\mu$ ($q'^\mu$) is the initial (final) momentum of the second particle, $\bar{K}^\mu =(p'^\mu - p^\mu)/\hbar=(q'^\mu - q^\mu)/\hbar$ is the momentum disparities between the final and initial states of each particle, and $\bar{\mathcal{A}}^{(0)}$ is the leading order scattering amplitude. 
	This expression effectively captures the momentum transfer involved in the scattering process.
	
	In this formalism, it is essential to consider the interplay of various length scales. In scattering involving massive particles, three critical length scales come into play:
	The Compton wavelength of the particles, characterized as $\ell_C^{(i)} = \hbar/m_i$ (where $m_i$ is the mass of particle $i$, $\hbar$ is the reduced planck constant, and $i=1,2$ corresponding to particle 1 and 2), serves as an indicator of the intrinsic quantum behavior exhibited by the particles.
	The intrinsic spread of the particles' wave packets, represented as $\ell_w^{(i)}$, quantifies our ability to precisely determine the particles' positions. It essentially serves as an intrinsic measure of the wavefunction's spatial distribution.
	The scattering length, denoted as $\ell_s$, is a product of the strength of interactions between particles and plays an important role in defining the scattering process's characteristics.
	
	Achieving an accurate description of the scattering process as that of point-like particles necessitates the fulfillment of specific conditions:
	$\ell_C^{(i)}\ll \ell_w^{(i)}$: This condition ensures that the intrinsic quantum behavior of particles does not dominate their spatial localization.
	$\ell_w^{(i)}\ll \ell_s$: This condition guarantees that the wave packet sizes remain significantly smaller than the characteristic length scale of the scattering process. It preserves the effective point-like nature of the particles. Balancing these conditions is crucial. If the wave packet spreads are too small, quantum effects become prominent, rendering the classical description inaccurate. Conversely, if the wave packet spreads are excessively large, particles lose their point-like character, leading to an imprecise classical trajectory description.


	\subsection{Real two-body scattering function at the first post-Minkowskian approximation}

\begin{figure}[t]
		\includegraphics[width=.35\textwidth]{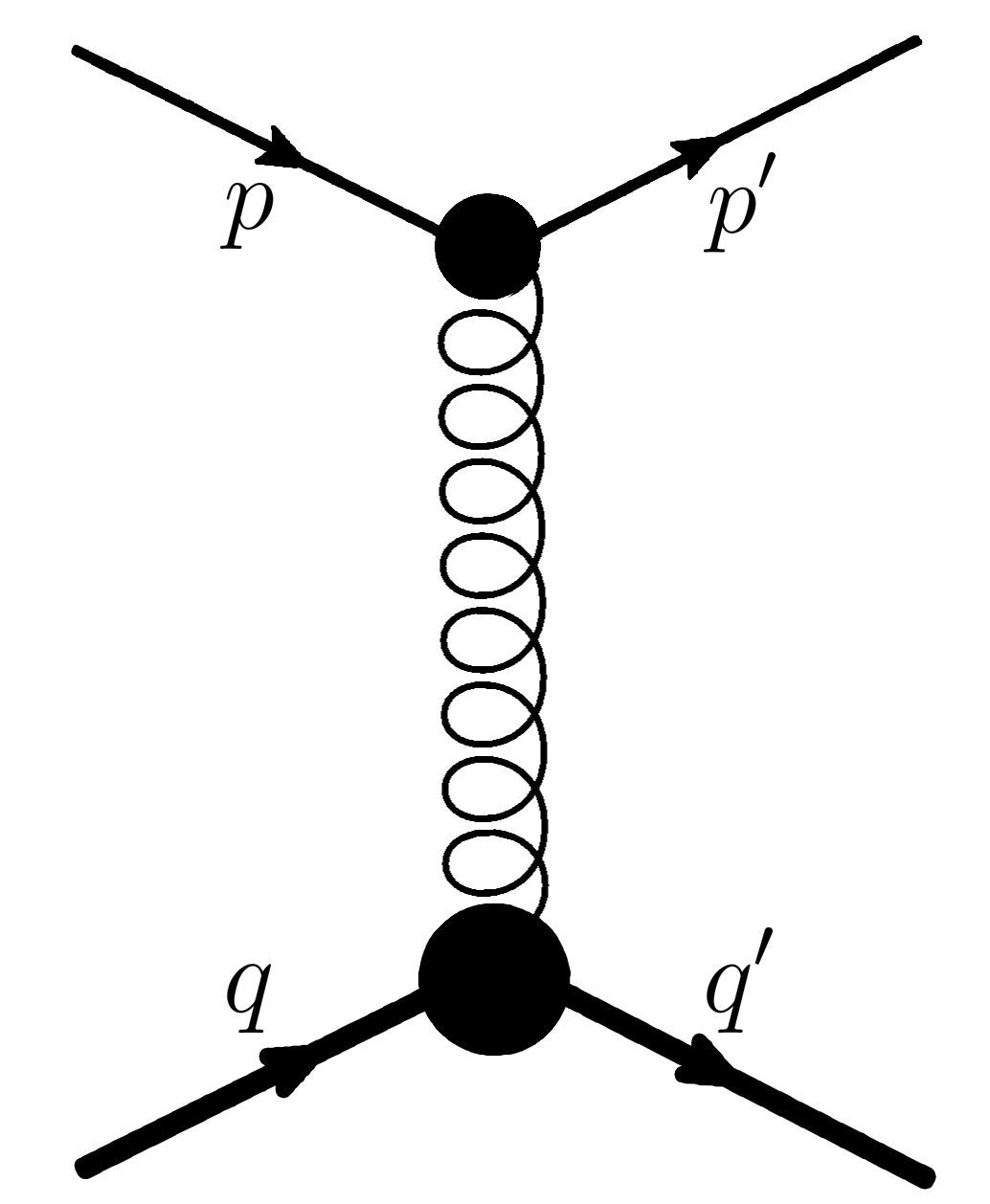}\centering
		\caption{Feynman diagram of the gravitational interaction between two massive objects (scalar fields with solid lines) via the graviton. $p$ ($q$) and $p'$ ($q'$) are the initial and final momenta of the system (environment) object, respectively.}
		\label{Feyn1}
	\end{figure}

	The gravitational impulse has also been evaluated in  \cite{Damour:2016gwp,Damour:2017zjx,Vines:2017hyw} using a relativistic gravitational two-body scattering approach. The new derivation discussed in \cite{Damour:2016gwp,Damour:2017zjx} is intended to show the connection between the classical scattering function  and the quantum scattering two-body amplitude. The author suggests that the new approach provides a clearer understanding of the physical principles that govern two-body scattering, the process depicted in Fig. \ref{Feyn1}.
	
	The evolution of the 4-momentum of particle 1, $p_{\mu}$, as it moves along its worldline $x^\mu_1(s_1)$ (with $s_1$ as the proper time) is given by
	\beq
	\frac{dp_{\mu}}{d\sigma_1}=\frac{1}{2}\partial_\mu g_{\alpha\beta}(x_1)
	p^\alpha p^\beta~,
	\eeq
	where $g_{\alpha\beta}(x_1)$ is the metric tensor at the particle's position $x_1$, and $\sigma_1=s_1/m_1$ is the rescaled proper time. 
	Integrating this equation with respect to $\sigma_1$ yields the total change $\Delta p_{\mu}$ in the particle's asymptotic 4-momentum, which is the difference between its initial and final momenta as it moves from $-\infty$ to $+\infty$ along its worldline
	\beq \label{Dp1}
	\Delta p_{\mu}=\int_{-\infty}^\infty d\sigma_1 \frac{1}{2} p^\alpha p^\beta \partial_\mu h_{\alpha\beta}(x_1)~,
	\eeq
	in which $h_{\alpha\beta}\equiv g_{\alpha\beta}-\eta_{\alpha\beta}$ with $\eta_{\alpha\beta}$ as the Minkowski metric. The expression \eqref{Dp1} gives the total change in 4-momentum of particle 1 due to the gravitational interaction with particle 2, at linear order in $G$. Let us now discuss the connection between this result and the Feynman diagram in Fig. \ref{Feyn1}.
	At linearized order in the gravitational constant $G$, the momentum $p^\alpha$ of particle 1 on the right-hand side of equation \eqref{Dp1} can be replaced by the constant incoming four-momentum of the particle. Moreover, the metric perturbation $h_{\alpha\beta}$ can be replaced by the solution of the linearized Einstein equations generated by the stress-energy tensor of particle 2. By substituting the expressions for $p^\alpha$ and $h_{\alpha\beta}$ induced by particle $2$ into equation \eqref{Dp1}, we obtain a relationship between the change in the momentum of particle $1$. 
	
	In this context, a metric perturbation is generated by the mass-energy distribution of particle 2 at the second order in perturbation theory. The Fourier transform of this perturbation is obtained in terms of the Fourier transform of the mass-energy distribution of particle 2 and the gravitational propagator
	\begin{eqnarray}\label{D1}
		D^{\alpha\beta\alpha'\beta'}(x) = \int \frac{d^4 K}{(2\pi)^4}D^{\alpha\beta\alpha'\beta'}(K)e^{-iK\cdot x}~,
	\end{eqnarray}
	in the harmonic (de Donder) gauge \cite{Bartolo:2018igk} with
	\begin{eqnarray}
		D^{\alpha\beta\alpha'\beta'}(K)= \frac{1}{K^2}\left(  \eta^ {\alpha \alpha' } \eta^ {\beta \beta' } + \eta^ {\alpha \beta} \eta^ {\beta \alpha' } 
		-\eta^ {\alpha\beta } \eta^ { \alpha' \beta' } 
		\right )~,
	\end{eqnarray}
	and $K^2 \equiv \eta^{\mu \nu} K_{\mu} K_{\nu}$ as the momentum transfer. Now, the expression for $\Delta p_{\mu}$ can be understood as follows: the stress-energy tensor of particle 2 acts as a source of gravitational waves, which propagate in spacetime with the gravitational propagator $D_{\mu\nu\alpha\beta}(K)$. The gravitational waves generated by particle $2$ then interact with particle $1$ and induces $\Delta p_{\mu}$ as the following form
	\bea \label{Dp2}
	\Delta p_{ \mu}= 8 \pi G \int  \frac{d^4 K}{(2\pi)^4} i K_{\mu} p^{\alpha} p^{\beta} D_{\alpha\beta\alpha'\beta'}(K) \,q^{\alpha'} q^{\beta'}
	\int d \sigma_1 \int d \sigma_2   e^{-i K.(x_1(\sigma_1) - x_2(\sigma_2))}\,.
	\eea
	This expression can be compared to the Feynman diagram depicted in Fig.~\ref{Feyn1} for scattering of two massive particles via graviton. In Feynman diagram language, the propagator $D_{\mu\nu\alpha\beta}(K)$ represents the exchange of a virtual graviton between the two particles, while two matter-gravity vertices $p^{\mu} p^{\nu}$
	and $q^{\alpha} q^{\beta} $ (computed in the approximation $p'\approx p$, $q'\approx q$)
	represents the coupling of the graviton to the mass-energy distribution of one particles. In Appendix~\ref{appendix:KMOCDAMOUR}, we have demonstrated that the results from the approach of two-body scattering function at the first post-Minkowskian approximation (Eq.~\eqref{Dp2}) and the result of the KMOC formalism (Eq.~\eqref{eq:KMOC}) are identical.

	\section{Gravitational impulse of two massive objects using QBE}\label{sec:massive}

	In this section, we describe the QBE for the gravitational scattering of two massive spinless particles via the exchange of virtual gravitons. In the case of the scattering of two massive objects, it is not clear which one should be considered the system and which one should be considered the environment. Furthermore, both massive objects are interacting with each other, so it is not straightforward to separate them into distinct system and environment components.
	That being said, we can still use the language of quantum systems and environments to describe the scattering process in a more general sense. We can think of one incoming massive objects as a quantum system that interact with another massive object that behaves as an environment.
	
	The scattering of two massive objects can be described using quantum field theory, where the gravitational interaction between two scalar fields is mediated by gravitons. The calculation of the scattering amplitude for this process can be represented by Feynman diagram of Fig.~\ref{Feyn1} involving the exchange of virtual gravitons between two scalar fields. In quantum field theory, the plane wave basis and the Gaussian basis are two commonly used methods for describing the quantum states of particles. 
	
	The plane wave basis is a mathematical representation of a quantum state in terms of a Fourier series of plane waves. In this basis, a quantum state is described by a superposition of plane waves, each with a different momentum and energy. Despite many advantage,  the plane wave basis does not provide a good description of particles that are localized in space, and it can be difficult to calculate observables such as position and momentum in this basis.
	
	The Gaussian basis, on the other hand, is a representation of a quantum state in terms of a superposition of Gaussian wave packets, each with a different center and spread. This basis is often used because it provides a good description of particles that are localized in space. When describing macroscopic particles in quantum field theory, the Gaussian basis is generally preferred over the plane wave basis. This is because macroscopic particles are typically localized in space and have a well-defined position, and the Gaussian basis provides a good description of such particles~\cite{Ishikawa:2018koj}.
	
	\subsection{Quantum Boltzmann equation}
	The classical Boltzmann equation already describes the statistical evolution of a thermodynamic system away from equilibrium. In a broader sense, it refers to any kinetic equation governing the time evolution of macroscopic quantities—such as energy, charge, or particle number—in a thermodynamic setting~\cite{Cercignani2012boltzmann}. To extend this framework to quantum systems, the classical Boltzmann equation is generalized to operate on the density matrix that describes the system’s quantum state. The formalism for this extended Boltzmann equation was first formulated to model neutrino flavor mixing in the presence of medium-induced interactions~\cite{Sigl1993general}. The technique was later generalized to give a QBE for any particular density matrix~\cite{kosowsky1995cosmic}. The advantage of this approach is that it treats particles in a general quantum manner described by a Boltzmann equation. It also gives a systematic perturbative expansion in the relevant small quantities and thus provides the framework for an investigation of higher order effects. Moreover, it allows the study of the mesoscopic behavior of an open quantum system by incorporating the microscopic interactions between the system and its environment, using standard techniques of quantum field theory. This framework is now being used for the time evolution of the intensity and polarization of cosmic microwave background (CMB) photons, detecting anisotropic gravitational waves, exploring CMB V-modes, and investigating the decoherence effects of squeezed gravitational waves~\cite{kosowsky1995cosmic,Bartolo:2018igk,Bartolo:2019eac,Hoseinpour:2020hic,Sharifian2024open}. 
	
	The time evolution of the system's number operator $\hat{\mathcal{N}}^\chi(\k)=\hat{a}^\dag(\k)a(\k)$, in the Heisenberg picture is 
	\begin{eqnarray}\label{eq:Heisenberg}
		\frac{d}{d t} \hat{\mathcal{N}}(\k)=i\left[\hat{H}, \hat{\mathcal{N}}(\k)\right]
	\end{eqnarray}
	where $\hat{H}$ is the full Hamiltonian. We decompose $\hat{H}$ into a free field contribution and an interaction term
	\begin{eqnarray}
		\hat{H}=\hat{H}_\textrm{free}+\hat{H}_\text{int}^\text{full}
	\end{eqnarray}
	with $\hat{H}_\text{int}^\text{full}$ a functional of the full fields. Expressing the right hand side of Eq.~\eqref{eq:Heisenberg} perturbatively in $\hat{H}_\text{int}$ leads to the QBE. We make the usual assumption of scattering theory that in a given interaction the fields begin as free fields and end as other free fields, and the interactions are isolated from each other. Consider the evolution of an operator through a single interaction beginning at $t=0$. Before this time, the fields can be taken as free to a good approximation. At $t=0$ the interaction Hamiltonian begins to turn on, and the interaction finishes at some later time, after which the fields can be taken as free once again. The time evolution of a general operator $\hat{\eta}^G$ to first order in the interaction Hamiltonian can be written as~\cite{Sigl1993general,kosowsky1995cosmic,Zarei:2021dpb}
	\begin{eqnarray}\label{eq:operatortime}
		\hat{\eta}^G(t)=\hat{\eta}(t)+i \int_0^t d t^{\prime}\left[\hat{H}_\text{int}\left(t-t^{\prime}\right), \hat{\eta}(t)\right]
	\end{eqnarray}
	where $\hat{\eta}(t)$ is the free-field operator satisfying the initial condition $\hat{\eta}(0)=\hat{\eta}^G(0)$, and $\hat{H}_\text{int}$ denotes the interaction Hamiltonian as a functional of the free fields. Eq.~\eqref{eq:operatortime} can be proven by taking the time derivative of both sides 
	\begin{eqnarray}
		\frac{d}{d t} \hat{\eta}^G(t)=\frac{d}{d t} \hat{\eta}(t)+i\left[\hat{H}_\text{int}(0), \hat{\eta}(t)\right]+i \int_0^t d t^{\prime} \frac{d}{d t}\left[\hat{H}_\text{int}\left(t-t^{\prime}\right), \hat{\eta}(t)\right]
	\end{eqnarray}
	The derivatives in the first and third terms can be replaced using the Heisenberg equation, but since they involve only free field operators, they evolve according to the free Hamiltonian $\hat{H}_\textrm{free}(t) = \hat{H}_\textrm{free}(0)$. Therefore, Eq.~\eqref{eq:operatortime} reduces to
	\begin{eqnarray}
		\frac{d}{d t} \hat{\eta}^G(t)=i\left[\hat{H}_\text{int}(0), \hat{\eta}(t)\right]+i\left[\hat{H}_\textrm{free}(0), \hat{\eta}(t)\right]
	\end{eqnarray}
	which, to first order in $\hat{H}_\text{int}$, reproduces the Heisenberg equation for $\hat{\eta}$.
	
	Using this result, the time evolution of $\hat{\mathcal{N}}$ can be expressed in terms of free-field operators. Substituting Eq.\eqref{eq:operatortime} into the commutator on the right-hand side of Eq.\eqref{eq:Heisenberg} gives
	\begin{eqnarray}\label{eq:numberev}
		\frac{d}{d t} \hat{\mathcal{N}}(\mathbf{k})=i\left[\hat{H}_\text{int}(t), \hat{\mathcal{N}}(\mathbf{k})\right]-\int_0^t d t^{\prime}\left[\hat{H}_\text{int}\left(t-t^{\prime}\right),\left[\hat{H}_\text{int}(t), \hat{\mathcal{N}}(\mathbf{k})\right]\right]
	\end{eqnarray}
	The integral on the right side can be made simpler and more useful by the so-called Born-Markov approximation. Under the Born approximation, the system interacts only weakly with its environment~\cite{Zarei:2021dpb}. With this context, the time step $t$ in Eq.~\eqref{eq:numberev} can be chosen large compared with a single collision and small compared to the time scale for density matrix evolution. After extending the time integral to infinity and taking the expectation value of both sides, we find  the time evolution of the number operator associated with the system’s degrees of freedom ~\cite{kosowsky1995cosmic,Zarei:2021dpb,Bartolo:2018igk,Bartolo:2019eac,Hoseinpour:2020hic,Sharifian2024open}
	\begin{eqnarray}\label{eq:numberevolution}
		\frac{d}{dt}\mean{\hat{\mathcal{N}}^\chi(\k)}=i\left < \left [\hat{H}_{\textrm{int}}(t),\hat{\mathcal{N}}^\chi(\k)\right ] \right > 
		-\frac{1}{2}\int_{-\infty}^\infty dt'
		\left < \left [\hat{H}_{\textrm{int}}(t),\left[\hat{H}_{\textrm{int}}(t'),\hat{\mathcal{N}}^\chi(\k)\right ] \right ]\right >~,\nonumber\\
	\end{eqnarray}
	where $\hat{H}_\text{int}(t)$ is the effective interaction Hamiltonian associated with the specific process of the system. The density operator of the system is defined by
	\begin{eqnarray}
		\hat{\rho}^\chi=\int \frac{d^3 \p'}{(2\pi)^3} \rho^\chi(\p')\hat{a}^\dag(\p')a(\p')~.
	\end{eqnarray}
	The expectation value of the number operator is connected to the density operator as
	\begin{eqnarray} \label{eq:numberexpectation}
		\mean{\hat{\mathcal{N}}(\k)}&=&\tr\l [\hat{\rho} \hat{\mathcal{N}}(\k) \r]=\int \frac{d^3 \p}{(2\pi)^3}  \bra{\p}  \hat{\rho}\hat{\mathcal{N}}(\k) \ket{\p} 
		\nonumber \\
		&=& 
		\int \frac{d^3 \p}{(2\pi)^3}   \frac{d^3 \p'}{(2\pi)^3} \rho(\p')\frac{1}{(p^0)^2}
		\bra{0} \hat{a}(\p)\hat{a}^\dag(\p')a(\p') \hat{a}^\dag(\k)a(\k) \hat{a}^\dag(\p)  \ket{0} 
		\nonumber \\
		&=&
		\int \frac{d^3 \p}{(2\pi)^3}   \frac{d^3 \p'}{(2\pi)^3} \rho(\p')\frac{1}{(p^0)^2}
		(2\pi)^9 k^0 p^0 p^0 \delta^3(\k-\p)\delta^3(\p'-\k)\delta^3(\p-\p')
		\nonumber \\
		&=&
		(2\pi)^3k^0 \delta^3(0) \rho(\k)~.
	\end{eqnarray}
	Therefore, by replacing the expectation value of Eq.~\eqref{eq:numberexpectation} in the time evolution of Eq.~\eqref{eq:numberevolution}, the QBE of the density matrix is derived
	\begin{eqnarray}\label{QBE00}
		(2\pi)^3\delta^3(0)k^0 \frac{d}{dt}\rho^\chi(\k)=i\left < \left [\hat{H}_{\textrm{int}}(t),\hat{\mathcal{N}}^\chi(\k)\right ] \right > 
		-\frac{1}{2}\int_{-\infty}^\infty dt'
		\left < \left [\hat{H}_{\textrm{int}}(t),\left[\hat{H}_{\textrm{int}}(t'),\hat{\mathcal{N}}^\chi(\k)\right ] \right ]\right >~.\nonumber\\
	\end{eqnarray}
	The first term on the right-hand side of Eq.\eqref{QBE00} is known as the forward scattering term while the second term is typically referred to as the collision term\cite{kosowsky1995cosmic, Zarei:2021dpb, Bartolo:2018igk}. Although the latter can include forward scattering effects, these effects vanish for uncorrelated target species~\cite{Sigl1993general}. In the QBE formulation, there is no need to impose the secular approximation and as shown in Refs.~\cite{kosowsky1995cosmic,Bartolo:2018igk,Bartolo:2019eac,Hoseinpour:2020hic,Sharifian2024open}, for the class of physical processes described by the interaction Hamiltonian, energy conservation emerges naturally once the integration over the microscopic time variable is carried out in the interaction picture. 
	
	The applicability of the Quantum Boltzmann Equation (QBE) is not restricted to situations involving large ensembles of particles. 
	Formally, the QBE comes from the open quantum system approach where the Hilbert space is split into a system and an environment so that the environment's degrees of freedom are traced out to yield an effective time evolution for the system's reduced density matrix. This procedure is neutral to the number of environmental degrees of freedom and the environment may consist of many particles, a medium, or (as in our case) even a single quantum object. In this work, we consider massive object $\chi$ as the system and the other massive object $\varphi$ as the environment that are interacting via graviton as depicted in Fig.~\ref{Feyn1}. Therefore, their interaction can be given by the second order S-matrix operator
	\begin{eqnarray}
		\hat{S}^{(2)} = -i \int_{-\infty}^{\infty} dt \, \hat{H}_\text{int}(t)~.
	\end{eqnarray}
	Until now, applications of the QBE to the evolution of photon, graviton and fermion systems have assumed that all interacting particles—both in the system and the environment—are fundamental point-like particles. In contrast, here we extend the QBE to incorporate macroscopic particles by modeling them as wave packets.

	\subsection{Effective interaction Hamiltonian}
	
	Here, we describe the gravitational scattering of two massive objects with scalar fields $\chi$ and $\varphi$. The interaction Hamiltonian density between scalar fields and graviton is given by
	\begin{eqnarray}\label{eq:Hdenscalar}
		\hat{\mathcal{H}}_{\chi\varphi h}=\frac{\kappa}{2}\, \hat{h}^{\mu\nu}(x)\partial_\mu\hat{\chi}(x) \partial_\nu \hat{\varphi}(x)~,
	\end{eqnarray}
	in which $\hat{h}^{\mu\nu}(x)$ is the mediating graviton field. In Fig.~\ref{Feyn1} we have plotted the Feynman diagram
	associated with the $ \chi$-$\varphi$ scattering process. We consider the particle $1$ as a quantum system that is described by
	\begin{eqnarray} \label{chi1}
		\hat{\chi}(x)= \hat{\chi}^+(x)+\hat{\chi}^-(x)=\int \frac{d^3 p}{(2\pi)^3p^0}\left [ \hat{a}(\p)e^{-ip\cdot x}+ \hat{a}^\dag(\p)e^{ip\cdot x} \right ] ~,
	\end{eqnarray}
	and the particle $2$ is the environment that can be represented as
	\begin{eqnarray} \label{varphi1}
		\hat{\varphi}(x)=\hat{\varphi}^+(x)+\hat{\varphi}^-(x)=\int \frac{d^3 q}{(2\pi)^3q^0} \left [\hat{b}(\q)e^{-iq\cdot x}+ \hat{b}^\dag(\q)e^{iq\cdot x} \right ] ~,
	\end{eqnarray}
	where $\hat{a}$ ($\hat{b}$) and $\hat{a}^\dag$ ($\hat{b}^\dag$) are the annihilation and creation operators of particles 1 (2) respectively, obeying the following canonical commutation
	relation
	\begin{eqnarray}
		\left[\hat{a}(\p),\hat{a}^\dag(\p')\right]=(2\pi)^3p^0 \delta^3(\p-\p')~,
	\end{eqnarray}
	and
	\begin{eqnarray}
		\left[\hat{b}(\q),\hat{b}^\dag(\q')\right]= (2\pi)^3q^0 \delta^3(\q-\q')~.
	\end{eqnarray}
	Also, the following expressions hold for the expectation value of the scalar fields' operators~\cite{Zarei:2021dpb}
	\begin{eqnarray}
		\left < \hat{a}^\dag(\p')\hat{a}(\p) \right > &=&p^0 (2\pi)^3\delta^3(\p-\p')\rho^\chi(\p)~,\label{eq:expda}
	\end{eqnarray}
	and
	\begin{eqnarray}
		\left < \hat{b}^\dag(\q')\hat{b}(\q) \right > &=&q^0 (2\pi)^3 \delta^3(\q-\q')\rho^\varphi(\q)~,\label{eq:expdb}
	\end{eqnarray}
	where $\rho^\chi(\p)$ and $\rho^\varphi(\q)$ characterize the density matrix of $\chi$ and $\varphi$ fields in the momentum space, respectively.  
	
	In QBE, the effective interaction Hamiltonian $\hat{H}_{\textrm{int}}$ is expressed in terms of the S matrix elements to represent the scattering process \cite{kosowsky1995cosmic,Zarei:2021dpb}. Here, we present $\hat{H}_{\textrm{int}}$ for the two-particle scattering process as follows
	\begin{eqnarray} \label{Hint1}
		\hat{H}_{\textrm{int}}(t)&=&\frac{\kappa^2}{4}\int d^3x d^4 x'd \tau d\tau'  \partial_\alpha\hat{\chi}^-(x) \partial_\beta\hat{\chi}^+(x) D^{\alpha\beta\alpha'\beta'}(x-x') \partial_\alpha' \hat{\varphi}^-(x') \partial_\beta' \hat{\varphi}^+(x')
		\nonumber \\ & &  \:\:\:\:\:\:\:\:\:\: \times \,
		V\delta^4(x-x(\tau))V'\delta^4(x'-x'(\tau'))~,
	\end{eqnarray}
	where $V$ and $V'$ are the considered volumes, $\tau$ and $\tau'$ are the proper time of two objects, and $\delta^4(x-x(\tau))$ is defined by
	\begin{eqnarray}\label{eq:delta4}
		\int_{-\infty}^{\infty}d\tau \delta^{4}(x-x(\tau))
		=\frac{d\tau}{dt}\delta^3(\mathbf{x}-\mathbf{x}(t))
		=\frac{1}{\gamma}\delta^3(\mathbf{x}-\mathbf{x}(t))~,
	\end{eqnarray}
	in which $\gamma$ is the Lorentz factor and $t=x^0(\tau)$. Here, $\delta^4(x - x(\tau))$ is a four-dimensional delta function that localizes the particle's contribution to the stress-energy tensor at its worldline $x(\tau)$. The Fourier transform of the interaction Hamiltonian is obtained by substituting Eqs. \eqref{chi1}, \eqref{varphi1} and \eqref{D1} into Eq.\eqref{Hint1}. The resulting interaction Hamiltonian is
	\begin{eqnarray} \label{Hint2-2}
		\hat{H}_{\textrm{int}}(t)&=&\frac{\kappa^2}{4}\int d^3x d^4 x' d \tau d\tau' d\p d\p' d\q d\q' dK\, p_\alpha p'_\beta q_{\alpha'} q'_{\beta' } D^{\alpha\beta\alpha'\beta'}(K) 	e^{-i(K+p-p')\cdot x}e^{-i(q-q'-K)\cdot x'}
		\nonumber \\  & &~\times V\delta^4(x-x(\tau))V'\delta^4(x'-x'(\tau'))	\hat{a}^\dag(\p')\hat{a}(\p) \hat{b}^\dag(\q')\hat{b}(\q)
		~,
	\end{eqnarray}
	with the following abbreviations
	\begin{eqnarray}
		dK \equiv \frac{d^4 K}{(2\pi)^4}~,~~~~~~~~~~~~~~ d\p \equiv \frac{d^3 p}{(2\pi)^3p^0}~,~~~~~~~~~~~~~~d\q \equiv \frac{d^3 q}{(2\pi)^3q^0}~.
	\end{eqnarray}
	This expression represents the interaction between particles in terms of their scattering amplitudes.
	We can substitute \eqref{Hint2-2} into the QBE to obtain the behavior of the density matrix.

	\subsection{Gaussian wave packets in the time evolution of the momentum operator}
	In the gravitational interaction between two massive particles, the number density of particles remains constant. In this type of interaction, we are interested in changes to the momentum operator associated with one of the particles, which we assume to be the system. When dealing with scalar fields, the momentum operator can be expressed using creation and annihilation operators as
	\begin{equation}
		\label{kmu}
		\hat{k}^\mu= \int \frac{d^3 k'}{(2\pi)^3}\frac{1}{k'^0}k'^\mu \hat{a}^\dagger_{\mathbf{k}'} \hat{a}_{\mathbf{k}'}~,
	\end{equation}
	where its expectation value is given by
	\begin{eqnarray}\label{eq:expected}
		\bar{k}^\mu \equiv\mean{\hat{k}^\mu}=\tr\l [\hat{\rho}\, \hat{k}^\mu \r]~,
	\end{eqnarray}
	where $\hat{\rho}$ is the reduced density matrix of the system. Therefore, we can use the QBE given in Eq.~\eqref{QBE00} to determine the time evolution of the momentum operator
	\begin{eqnarray} \label{QBEkmu0}
		\frac{d}{dt}\,\bar{k}^\mu=i\left < \left [\hat{H}_{\textrm{int}}(t),\hat{k}^\mu\right ] \right > 
		-\frac{1}{2}\int_{-\infty}^\infty dt' 
		\left < \left [\hat{H}_{\textrm{int}}(t'),\left[\hat{H}_{\textrm{int}}(t ),\hat{k}^\mu\right ] \right ]\right >~.
	\end{eqnarray}
	In the following, we will focus exclusively on the forward scattering term, which will serve as the foundation for calculating the impulse.
	
	However, it is important to note that there are distinct differences in the forward scattering term between interacting particles that are localized and those that are not. Therefore, we will first address this distinction. Subsequently, we will proceed to calculate the forward scattering term for gravitational scattering of two massive objects.

	In the standard theoretical description of the scattering process, the incoming and outgoing particles are described by a plane waves. Such an approach is valid for a number of scattering problems in which the distances necessary to calculate the corresponding cross sections are significantly smaller than typical particle sizes.
	However, there are important examples in which macroscopically large impact parameters gave an essential contribution to the cross section. These impact parameters may be much larger than the transverse sizes of the colliding bunches. In that case, the standard calculations have to be essentially modified. One solution is to use Gaussian wave packet to describe the scattering of macroscopic particles on macroscopic targets \cite{Karlovets:2017xgg,Karlovets:2016jrd,Karlovets:2016dva,Karlovets:2017gzk}. Gaussian wave packets can be used to characterize the quantum state of macroscopic particles that are spatially large and do not belong to the category of elementary particles in quantum field theory \cite{kosowsky1995cosmic}. These states represent a specific class of quantum states with a Gaussian probability distribution in phase space.
	
	To deal with the forward scattering of macroscopic particles and consider the wave packet states it is necessary to generalize the forward scattering term. As a result, the original term appeared above can no longer be referred to as the forward scattering term. Instead, it is now referred to as the quasi-forward scattering term. The impulse is then calculated based on this term.

	Here, we explicitly calculate the gravitational forward scattering of two massive particles and show how a finite width of the particle world tubes gives rise to an uncertainty in momentum in each vertices. For this reason, we replace the $\delta^4(x-x(\tau))$ functions in the interaction Hamiltonian of Eq.~\eqref{Hint1} with the smeared delta function \cite{Breuer:2002pc},
	\begin{eqnarray}\label{eq:delta4bar}
		\delta^4_{\sigma_\chi}(x-\bar{x}(\tau))=\delta(x^0-\bar{x}^0)\frac{1}{(2\pi \sigma^2_\chi)^{3/2}}\exp\l [- \frac{(\x-\bar{\x})^2}{2\sigma_\chi^2}\r]~,
	\end{eqnarray}
	which has a Gaussian distribution with width $\sigma_\chi$. The same happens for the $\delta^4(x'-x'(\tau'))$ with variance, $\sigma_\varphi$. To clarify the role of wave packets, we have calculated physical quantities such as current and energy-momentum tensor in Appendix~\ref{appendix:currentandstress}. Substituting Eq.\eqref{Hint2-2}  into the forward scattering term of Eq.\eqref{QBEkmu0} with considering the smeared delta function, and then taking integration over $x$ and $x'$ leads to
	\begin{eqnarray} \label{Deltak1}
		\Delta \bar{k}^{(1)\mu} &=&
		i\frac{\kappa^2}{4}\int 
		d\tau d\tau' 
		d\p d\p'  d\q d\q'
		dK  p_{\alpha} p'_{\beta} \bar{q}_{\alpha'}\bar{ q}'_{\beta'}   
		D^{\alpha\beta\alpha'\beta'}(K) 
		e^{-i(K+p-p')\cdot x(\tau)}e^{-i(q-q'-K)\cdot x(\tau')}
		\nonumber \\  &   \times & \;
		e^{-\sigma_\chi^2|\K+\p-\p'|^2/2}e^{-\sigma_\varphi^2|\K+\q'-\q|^2/2}V\,V'\nonumber\\
		&   \times & \;	\l <  \hat{b}^\dag(\q')\hat{b}(\q)  \right >
		\left [ \l <\hat{a}^\dag(\p')\hat{a}(\p) \hat{k}^\mu  \right > -
		\l < \hat{k}^\mu  \hat{a}^\dag(\p')\hat{a}(\p)  \right >
		\r ]
		~.\nonumber\\	
	\end{eqnarray}
	We see that the finite widths $\sigma_\chi$ and $\sigma_\varphi$ of the current world tubes yield effective ultraviolet cutoffs in \eqref{Deltak1} and imply that the particle's momenta are not precisely defined.  
	The spreading of the wave packet leads to uncertainty in the momentum of the scattered particles. This uncertainty arises because the spread of the wave packet limits our ability to precisely determine the momentum of the scattered particles.
	
	\subsection{Gravitational impulse from forward scattering of Gaussian wave packets}
	In the forward scattering condition, once we integrate over momenta, it automatically yields $\p=\p'$, indicating that the particle's momentum remains unchanged. However, for macroscopic particles, the Gaussian expression, $\exp\l(\frac{-\sigma^2_\chi|\K+\p-\p'|^2}{2}\r)$, implies that the incoming particle's momentum will change, with $\abs{\Delta \p} = 1/\sigma_\chi$
	where $\sigma_\chi$ is the average size of the packet. We assume that this dispersion is small compared to the momentum $1/\sigma_\chi \ll |\p|$.

	If we calculate the expected values for plane wave fields with the equation \eqref{Deltak1}, the forward scattering term on the right side of \eqref{Deltak1} vanishes because 
	\begin{multline}\label{eq:expdif}
		\int 
		d\p   d\p' \l[\l <\hat{a}^\dag(\p')\hat{a}(\p) \hat{k}^\mu  \right > -
		\l < \hat{k}^\mu  \hat{a}^\dag(\p')\hat{a}(\p)  \right >\r]\\=\int 
		d\p   d\p' \l[(p^\mu-p'^\mu)
		p^0(2\pi)^3\delta^3(\p-\p')\rho^\chi(\p)\r]=0~.
	\end{multline}
	Here, due to the Gaussian expressions, $\p$ and $\q$ are dispersed around their mean value  with dispersion $\sigma_\chi$ and $\sigma_\varphi$ respectively. Therefore, after integrating over $\p'$ and $\q'$, we obtain
	\begin{eqnarray} \label{Deltak2}
		\Delta \bar{k}^{(1)\mu}
		&=&
		i\frac{\kappa^2}{4}\int 
		d\p   d\q 
		dK K_\mu k_{\alpha} k_{\beta} q_{\alpha'} q_{\beta'}   
		D^{\alpha\beta\alpha'\beta'}(K) 
		e^{-(\sigma_\chi^2+\sigma_\varphi^2)|\K|^2/2}
		V'n_\varphi(\x,\q)Vn_\chi(\x,\k)\nonumber\\
		&&\times\int d\sigma \int d\sigma'  e^{-iK\cdot (x(\sigma)-x(\sigma'))}
		~.
	\end{eqnarray}
	Here, the densities $\rho^\varphi(\q)$ and $\rho^\chi(\k)$ have been replaced by the corresponding number densities $n_\varphi(\x,\q)$ and $n_\chi(\x,\k)$, and we have used the parametrizations $\sigma=\tau/k^0$ and $\sigma'=\tau'/q^0$. We consider a Gaussian distribution for $n_\varphi(\x,\q)$ and $n_\chi(\x,\k)$ as follows~\cite{kosowsky1995cosmic}
	\begin{eqnarray}
		n_\chi(\x,\k)=n_\chi(\x) \l ( \frac{2\pi}{\sigma^2_\k} \r)^{3/2}\exp \l [- \frac{(\k-\bar{\k})^2}{2\sigma_\k^2}\r ]~,
	\end{eqnarray}
	and
	\begin{eqnarray}
		n_\varphi(\x,\q)=n_\varphi(\x) \l ( \frac{2\pi}{\sigma^2_\q} \r)^{3/2}\exp \l [- \frac{(\q-\bar{\q})^2}{2\sigma_\q^2}\r ]~,
	\end{eqnarray}
	where $\sigma_\k^2$ ($\sigma_\q^2$) and $\bar{\k}$ ($\bar{\q}$) are the distribution variance and average momentum of particle $\chi$ ($\varphi$) in the momentum space, respectively. Integrating over $\k$ and $\q$ gives
	\begin{eqnarray}
		\int d\k k_{\alpha} k_{\beta}  n_\chi(\x,\k)=\bar{k}_{\alpha} \bar{k}_{\beta}n_\chi(\x)~,
	\end{eqnarray}
	and
	\begin{eqnarray}
		\int d\q q_{\alpha'} q_{\beta'}n_\varphi(\x,\q)=\bar{q}_{\alpha'} \bar{q}_{\beta'}n_\varphi(\x)~,
	\end{eqnarray}
	with bars over the four momentum indicating its average value. For a single massive object we have $Vn_\chi(\x)=V'n_\varphi(\x)=1$ and therefore
	\begin{eqnarray} \label{deltkmu-5}
		\Delta \bar{k}^{(1)\mu}   = 
		8\pi G\int \frac{d^4K}{(2\pi)^4}
		iK^\mu \bar{k}_{\alpha} \bar{k}_{\beta} \bar{q}_{\alpha' }\bar{ q}_{\beta'}   
		D^{\alpha\beta \alpha'\beta'}(K) e^{-(\sigma_\chi^2+\sigma_\varphi^2)|\K|^2/2}
		\int d\sigma \int d\sigma'  e^{-iK\cdot (x(\sigma)-x(\sigma'))}
		~.\nonumber\\
	\end{eqnarray}
	In the situation where the Gaussian function can be taken as 1, this expression matches the result \eqref{Dp2}. This condition is fulfilled if
	\begin{eqnarray}\label{lambdag1}
		|\K| \ll \frac{1}{\ell_{w}}~,
	\end{eqnarray}
	where $\ell_w$ is the wave packet length scale $\ell_w=\text{max}(\sigma_\chi,\sigma_\varphi)$.
	Our analysis treats the matter states as wave packet states. We see that the finite widths $\sigma_\chi$ and $\sigma_\varphi$ yield effective ultraviolet cutoff for $|\K|$ as given in Eq \eqref{deltkmu-5}. 
	We are also interested in exploring the limit where the matter currents can be treated as classical currents. This approximation can be justified under the following conditions. We assume that the wavelength of the graviton, $\lambda_g$, emitted by the matter current 1 is large compared to the Compton wavelength of the particle 1, $\lambda^{(1)}_{C}$
	\begin{eqnarray}\label{lambdag2}
		\lambda_g \gg \lambda^{(1)}_{C}~.
	\end{eqnarray}
	In the low energy region, we neglect the bremsstruhlung graviton radiation from the matter field 1 during the impulse process. 
	In an experiment of the type sketched in Fig.~\ref{Feyn1}, the gravitational impulse of passing particle 1 causes particle 1 to accelerate as it moves through the gravitational field generated by particle 2.
	This force results in a specific characteristic acceleration time denoted as $\tau_p$ or equivalently the scattering length scale $\ell_s \sim \tau_p$. We define $\tau_p$ as the reciprocal of the highest frequency present in the power spectrum of the gravitational force acting on particle 1. Due to the presence of this characteristic time, we have a natural upper limit $|\K|$ for the frequency spectrum of the gravitational force, which is of the order of
	\begin{eqnarray} 
		|\K|\sim \frac{1}{\ell_s}~.
	\end{eqnarray}
	Our requirement in Eq.~\eqref{lambdag2} thus takes the form
	\begin{eqnarray} \label{ellw2}
		\ell_w \gg \lambda^{(1)}_{C}~.
	\end{eqnarray}
	Combining the conditions \eqref{lambdag1} and \eqref{ellw2} gives rise to the Goldilocks’ zone \cite{Kosower:2018adc}
	\begin{eqnarray} \label{ellw3}
		\lambda^{(1)}_{C}\ll \ell_w \ll \ell_s~. 
	\end{eqnarray}
	A result of particular interest from a fundamental point of view is that when particles 1 and 2 are considered fundamental particles with variances close to zero and the wave packet approximation is replaced by a plane wave, the impulse in the leading order becomes zero, as can be verified in equation \eqref{deltkmu-5}.
	This result can be confirmed by observing that the gravitational forward scattering of two fundamental particles is always zero. However, as we have seen in our main result \eqref{deltkmu-5}, the forward scattering term leads to a non-zero impulse in the leading order only when the particles are macroscopic and due to the property of their wave function being a wave packet.
	
	\subsection{Second order term: collition term} 
	\label{sec:second}
	So far, we have discused the leading order tree-level or 1-graviton exchange terms ($\mathcal{O}(G)$). The next-to-leading order terms ($\mathcal{O}(G^2)$) typically include 1-loop diagrams and also higher-order tree diagrams involving multi-graviton exchange.
	Here, we will be focusing the higher order tree diagrams and discuss the meaning of such diagrams in the context of out extended QBE appraoch. 
	The contribution of second order term ($\mathcal{O}(G^2)$) in the gravitational impulse $\Delta \bar{k}^{(2)\mu}$ is given by the collision term (second term) in eq. \eqref{QBEkmu0}
	\begin{eqnarray} \label{QBEkmu-1-1}
		\text{Collisional contribution}=	-\frac{1}{2}\int_{-\infty}^\infty dt \,ds
		\left < \left [\hat{H}_{\textrm{int}}(s),\left[\hat{H}_{\textrm{int}}(t ),\hat{k}^\mu\right ] \right ]\right >~.
	\end{eqnarray}
	We will use the interaction Hamiltonian \eqref{Hint2-2} in \eqref{QBEkmu-1-1} to calculate the contribution of the collision term as
	\begin{eqnarray} \label{QBEkmu-1-2}
		\Delta \bar{k}^{(2)\mu} &=&		-\frac{\kappa^4}{32} \int_{-\infty}^\infty dt\,ds  \int   d^3x_1 d^4 x'_1 d \tau_1 d\tau'_1 d\p_1 d\p'_1 d\q_1 d\q'_1 dK_1 d^3x_2 d^4 x'_2 d \tau_2d\tau'_2 d\p_2 d\p'_2
		\nonumber \\  & &~\times
		d\q_2 d\q'_2 dK_2
		p_{1\alpha} p'_{1\beta} q_{1\alpha'} q'_{1\beta' }
		p_{2\alpha} p'_{2\beta} q_{2\alpha'} q'_{2\beta' } 
		D^{\alpha_1\beta_1\alpha'_1\beta'_1}(K_1)
		D^{\alpha_2\beta_2\alpha'_2\beta'_2}(K_2) 
		\nonumber \\  & &~\times
		e^{-i(K_1+p_1-p'_1)\cdot x_1}e^{-i(q_1-q'_1-K_1)\cdot x'_1}
		e^{-i(K_2+p_2-p'_2)\cdot x_2}e^{-i(q_2-q'_2-K_2)\cdot x'_2}
		\nonumber \\  & &~\times V\delta^4(x_1-x_1(\tau_1))V'\delta^4(x'_1-x'_1(\tau'_1))
		V\delta^4(x_2-x_2(\tau_2))V'\delta^4(x'_2-x'_2(\tau'_2))
		\nonumber \\  & &~\times
		\left < \left [\hat{a}^\dag(\p'_1)\hat{a}(\p_1) \hat{b}^\dag(\q'_1)\hat{b}(\q_1),\left[\hat{a}^\dag(\p'_2)\hat{a}(\p_2) \hat{b}^\dag(\q'_2)\hat{b}(\q_2),\hat{k}^\mu\right ] \right ]\right >~.
	\end{eqnarray}
	After ingeration over $x_1$, $x'_1$, $\p_1$,  $\p_2$, $\q_1$, and $\q_2$
	(the detailed derivation of  is presented in Appendix \eqref{C}) we finally get  following expresion for the $\Delta \bar{k}^{(2)\mu}$
	\begin{eqnarray}
		\Delta \bar{k}^{(2)\mu} &=&
		\frac{\kappa^4}{16}\int 
		d\sigma_2 d\sigma'_2
		dK_1 dK_2
		e^{-\sigma_\chi^2(|\K_1|^2+|\K_2|^2)/2}e^{-\sigma_\varphi^2(|\K_1|^2+|\K_2|^2)/2} 
		e^{-i(K_1+K_2)\cdot ( x_2(\sigma_2)-x_2(\sigma'_2))}
		\nonumber \\  &    \times & \;	(K_1^\mu-K_2^\mu) \,
		[ \bar{p}_{1\alpha_1} \bar{p}_{2\beta_1}
		D^{\alpha_1\beta_1\alpha'_1\beta'_1}(K_1)
		\bar{q}_{1\alpha'_1}\bar{q}_{2\beta'_1} ]\, [
		\bar{p}_{2\alpha_2} \bar{p}_{1\beta_2} 
		D^{\alpha_2\beta_2\alpha'_2\beta'_2}(K_2) 
		\bar{q}_{2\alpha'_2}\bar{q}_{1\beta'_2}]\,,\nonumber\\
	\end{eqnarray}
	which reproduces the same physical structure as second order impulse momentum of Ref.~\cite{Kosower:2018adc}. Using an appropriate redefinition of parameters (e.g., wave-packet widths, $q\rightarrow K_1+K_2$ and $w_1\rightarrow K_1-K_2 $), the two expressions coincide. This demonstrates that our QBE approach is fully consistent with the established order $\mathcal{O}(G)$ results of the KMOC formalism.

	In Ref.~\cite{Kosower:2018adc}, the authors compute loop corrections up to five loops, corresponding to specific diagrammatic topologies such as triangles, boxes, and cut boxes. Nevertheless, it is important to emphasize that within the QBE framework these corrections can, in principle, be obtained systematically—either by including loop corrections to the interaction Hamiltonian or by incorporating the forward term. In this manner, the approach can reproduce the same loop contributions as those reported in Ref.~\cite{Kosower:2018adc}.

	
	\section{Gravitational deflection of light by a massive object} 
	\label{sec:light}
	
	Another intriguing application of the derived formulas from the preceding section pertains to the gravitational bending of light caused by a massive object. This phenomenon can be studied by calculating the alteration in momentum of a narrow light beam that passes near a massive point-like particle with a non-zero impact parameter $b$.
	
	In \cite{Cristofoli2022waveforms}, the authors incorporated massless classical waves in the KMOC formalism as initial states. It is important to highlight that extending the application of the KMOC formalism to massless particles is unfeasible. For massless boson particles, the divergence of the particle's Compton wavelength renders it impossible to satisfy the Goldilock's condition. Treating photons as point-like particles is impossible, as demonstrated by the rigorous proofs of Newton and Wigner \cite{Newton1949localized} and Wightman \cite{Wightman1962on} that pinpointing the exact position of known massless particles in space is unattainable. Instead, a proper approach necessitates the use of coherent states. The authors in \cite{Cristofoli2022waveforms} delved into the fundamental aspects of coherent states, with a specific focus on the electromagnetic scenario. They describe a localized incoming beam of light by choosing the beam localized around its center. 
	
	In this section, we explore the interpretation of gravitational impulse of photons using QBE. 
	We employ the findings from \cite{Cristofoli2022waveforms} to describe the behavior of photons propagating within a gravitational field of a spherical mass.
	
	\begin{figure}[t]
		\vspace{-2cm}
		\includegraphics[width=.3\textwidth]{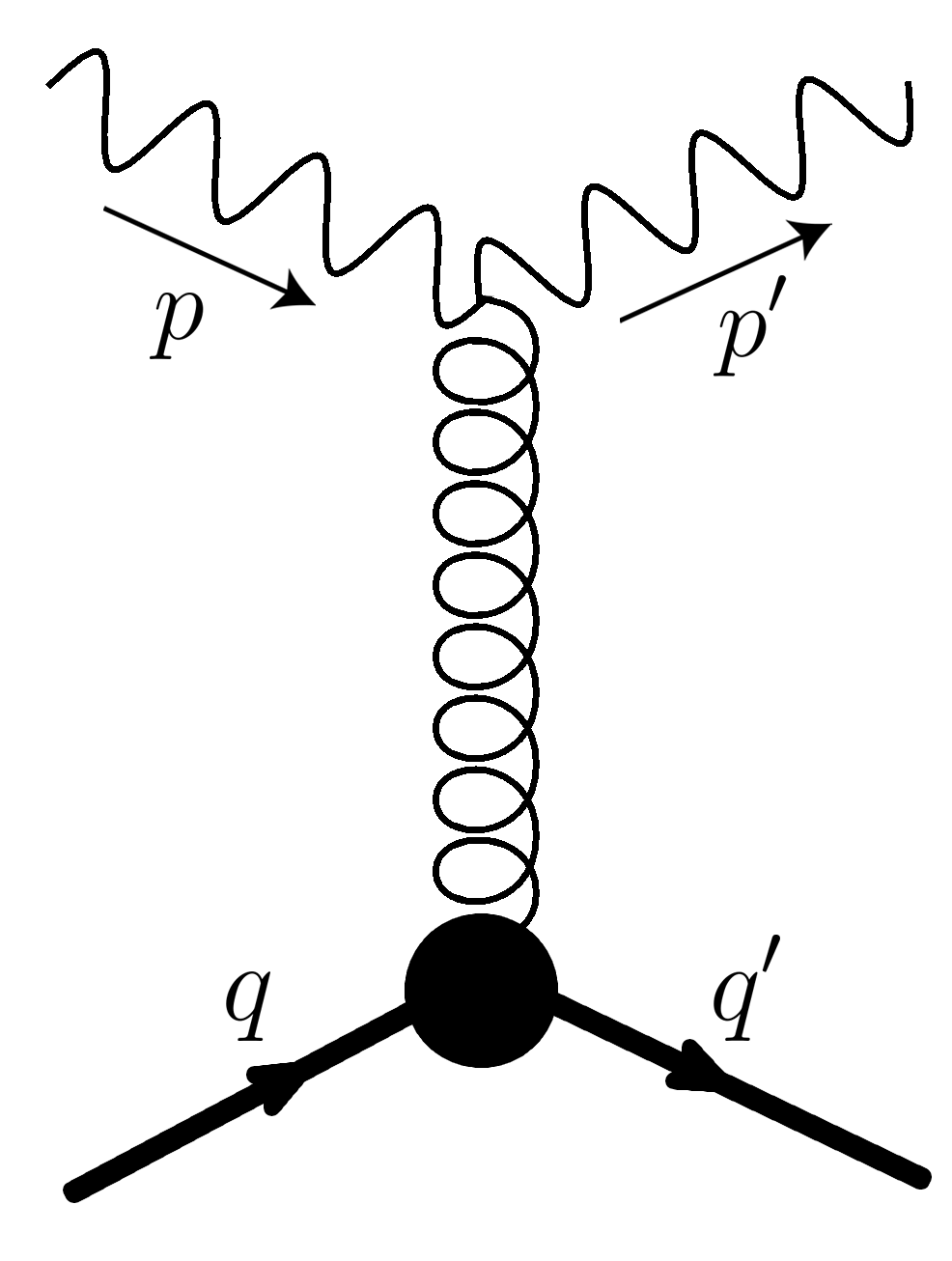}\centering
		
		\caption{Feynman diagram depicting the gravitational impulse of light due a massive object mediated by gravitons. $p$ ($q$) and $p'$ ($q'$) are the initial and final momenta of the photon (massive object), respectively.}
		\label{fig:scalar&photon}
	\end{figure}

	Initiating with the interaction Hamiltonian density of photons with gravitational field
	\begin{eqnarray}
		\hat{\mathcal{H}}_{\gamma\gamma h}=\frac{\kappa}{2}\, \hat{h}^{\mu\nu}(x)\hat{F}_\mu^{\;\lambda}(x) \hat{F}_{\lambda\nu}(x)~,
	\end{eqnarray}
	and using Eq.~\eqref{eq:Hdenscalar}, the effective Hamiltonian describing the interaction depicted by Feynman diagrams in Fig.~\ref{fig:scalar&photon} is
	\begin{eqnarray}
		\hat{H}_{\inte}&=&\frac{\kappa^2}{4} \int d^3 x d^4 x'd\tau d\tau' D^{\alpha\beta\gamma\delta}(x-x')\hat{F}_\alpha^{+~\lambda}(x)\hat{F}^-_{\lambda \beta}(x)\partial_\gamma \hat{\varphi}^+(x') \partial_\delta \hat{\varphi}^-(x')\nonumber\\
		&&\times V\delta^4(x-x(\tau)) V'\delta^4(x'-x'(\tau'))
		~.
	\end{eqnarray}
	where $\hat{F}_{\lambda\beta}(x)=\partial_\lambda \hat{A}_\beta(x)-\partial_\beta \hat{A}_\lambda(x)$ is the photon field strength tensor. The Fourier transform of the photon field $A^\alpha(x)$ is defined using the following convention
	\begin{eqnarray}
		\hat{A}^\alpha(x)=\hat{A}^{\alpha +}(x)+ \hat{A}^{\alpha -}(x)
		=\int d\p \sum_s \left [ \hat{c}_s(\p) \varepsilon_s^\alpha(\p) e^{-ip\cdot x}
		+ \hat{c}^\dag_s(\p) \varepsilon_s^{\alpha *}(\p) e^{ip\cdot x}\right ]
		~,
	\end{eqnarray}
	in which $\hat{c}_s(\p)$ and $\hat{c}^\dagger_s(\p)$ are the annihilation and creation operators satisfying the following commutation relation  
	\begin{eqnarray}
		\left[\hat{c}_s(\p),\hat{c}_{s'}^\dag(\p')\right]=(2\pi)^3p^0 \delta^3(\p-\p')\delta_{s,s'}~,
	\end{eqnarray}
	and $\varepsilon_s^\alpha(\p)$ are photon polarization 4-vectors with the polarization indexes $s=1,2$. The following expectation value for the photon number operator holds
	\begin{eqnarray}
		\left < \hat{c}^\dag(\p')\hat{c}(\p) \right > =p^0(2\pi)^3 \delta^3(\p-\p')\delta_{s,s'}\rho^\gamma(\p)~,
	\end{eqnarray}
	where $\rho^\gamma(\p)$ is the density matrix of the photon in the momentum space and has the following relation with its density in the position space $n_\gamma(\x)$
	\begin{eqnarray}\label{eq:photonrho}
		\int d\p\,p_\alpha p_\beta \rho^\gamma(\p)\delta_{s,s'}=\bar{p}_\alpha\bar{p}_\beta n_\gamma(\x)\delta_{s,s'}~,
	\end{eqnarray}
	with $\bar{p}_\alpha$ as the average value of the four momentum in a beam of photons.
	Using the Fourier transforms of Eq.~\eqref{D1} for the propagator and Eq.~\eqref{varphi1} for the scalar field, we have 
	\begin{eqnarray}
		\hat{H}_{\inte}(t)&=&\frac{\kappa^2}{4}\sum_{s,s'}\int d^3 x d^4 x' d\tau d\tau' d\p d\p' d\q d\q'dK \, D^{\alpha\beta\gamma\delta}(K)	e^{-ix\cdot(K+p-p')}e^{-ix'(\cdot(q-q'-K)}\nonumber\\
		&&\times~ V\delta^4(x-x(\tau))V'\delta^4(x'-x'(\tau'))
		\nonumber \\ &&\times~
		\left[ p_\alpha \varepsilon_{s}^\lambda(\p)- p^\lambda \varepsilon_{s\alpha}(\p) \right]
		\left[p'_\lambda \varepsilon_{s'\beta}(\p')- p'_\beta \varepsilon_{s'\lambda}(\p') \right]
		q_\gamma q'_\delta \hat{c}^\dag_{s'}(\p')\hat{c}_s(\p)\hat{b}^\dag(\q')\hat{b}(\q)	~.\nonumber\\
	\end{eqnarray}
	Therefore, the quantum Boltzmann equation in Eq.~\eqref{QBEkmu0} up to the forward scattering term for describing the time evolution of $\hat{k}^\mu$ leads to
	\begin{eqnarray}
		\Delta \bar{k}^{(1)\mu} &=&\frac{\kappa^2}{4}\sum_{s,s'}\int d^3 x d^4 x' d\tau d\tau' d\p d\p' d\q d\q'dK \, D^{\alpha\beta\gamma\delta}(K)	e^{-ix\cdot(K+p-p')}e^{-ix'\cdot(q-q'-K)}\nonumber\\
		&&\times~ V\delta^4(x-x(\tau))V'\delta^4(x'-x'(\tau'))\left[ p_\alpha \varepsilon_{s}^\lambda(\p)- p^\lambda \varepsilon_{s\alpha}(\p) \right]
		\left[p'_\lambda \varepsilon_{s'\beta}(\p')- p'_\beta \varepsilon_{s'\lambda}(\p') \right]
		q_\gamma q'_\delta\nonumber\\
		&&\times~\left< \hat{b}^\dag(\q')\hat{b}(\q) \right >\left<\left [\hat{c}^\dag_{s'}(\p')\hat{c}_s(\p),\hat{k}^\mu\right ]\right>~.\label{eq:dkph}
	\end{eqnarray}
	As in Section~\ref{sec:massive}, we can see that by considering plane wave fields the forward scattering term vanishes because the expectation value of the commutator term is
	\begin{multline}
		\int d\p d\p'\l[\braket{\hat{c}_{s'}^\dag(\p')\hat{c}_s(\p) \hat{k}^\alpha}-\braket{\hat{k}^\alpha \hat{c}_{s'}^\dag(\p')\hat{c}_s(\p)}\r]\\=\int d\p d\p'\l[(p^\alpha-p'^\alpha)\rho^\gamma(\mathbf{p}) p^0(2\pi)^{3}\delta^3(\mathbf{p}-\mathbf{p}')\delta_{s,s'}\r]=0~.
	\end{multline}
	Therefore, the delta functions in Eq.~\eqref{eq:dkph} should be replaced by the smeared delta functions described in Eq.~\eqref{eq:delta4bar}. With this replacement and integrating over $x$ and $x'$, Eq.~\eqref{eq:dkph} changes into
	\begin{eqnarray}
		\Delta \bar{k}^{(1)\mu}&=&i\mean{\left [
			\hat{H}_{\inte},\hat{k}^\mu \right ]}
		\nonumber \\ 
		&=&i\frac{\kappa^2}{4}\sum_{s,s'}\int d\tau d\tau' d\p d\p' d\q d\q'dK D^{\alpha\beta\gamma\delta}(K)e^{-\sigma_\gamma^2|\K+\p-\p'|^2/2}e^{-\sigma_\varphi^2|\K+\q'-\q|^2/2}
		\nonumber \\ &&\times~e^{-iK\cdot (x(\tau)-x'(\tau'))}
		\left[ p_\alpha \varepsilon_{s}^\lambda(\p)- p^\lambda \varepsilon_{s\alpha}(\p) \right]
		\left[p'_\lambda \varepsilon_{s'\beta}(\p')- p'_\beta \varepsilon_{s'\lambda}(\p') \right]
		q_\gamma q'_\delta 	\nonumber\\
		&&\times~V\,V'\,\left< \hat{b}^\dag(\q')\hat{b}(\q) \right >\left<\left [\hat{c}^\dag_{s'}(\p')\hat{c}_s(\p),\hat{k}^\mu\right ]\right>~,
	\end{eqnarray}
	and using the expectation values leads to
	\begin{eqnarray}
		\Delta \bar{k}^{(1)\mu}&=&i\frac{\kappa^2}{4}\sum_{s,s'}\int d\tau d\tau' d\p d\p' d\q d\q'dK D^{\alpha\beta\gamma\delta}(K)e^{-\sigma_\gamma^2|\K+\p-\p'|^2/2}e^{-\sigma_\varphi^2|\K+\q'-\q|^2/2}
		\nonumber \\ &&\times~
		\left[ p_\alpha \varepsilon_{s}^\lambda(\p)- p^\lambda \varepsilon_{s\alpha}(\p) \right]
		\left[p'_\lambda \varepsilon_{s'\beta}(\p')- p'_\beta \varepsilon_{s'\lambda}(\p') \right]
		q_\gamma q'_\delta 	q^0 (2\pi)^3 \delta^3(\q-\q')\rho^\varphi(\q)\nonumber\\&&\times~e^{-iK\cdot (x(\tau)-x'(\tau'))}V\,V' (p^\alpha-p'^\alpha)p^0(2\pi)^{3}\delta^3(\mathbf{p}-\mathbf{p}')\delta_{s,s'}\rho^\gamma(\mathbf{p}) \nonumber\\
		&=&i\frac{\kappa^2}{4}\sum_{s}\int d\tau d\tau' d\p dK D^{\alpha\beta\gamma\delta}(K)e^{-(\sigma_\gamma^2+\sigma_\varphi^2)|\K|^2/2}e^{-iK\cdot (x(\tau)-x'(\tau'))}
		\nonumber \\ &&\times~V\,V'\,
		\left[ p_\alpha \varepsilon_{s}^\lambda(\p)- p^\lambda \varepsilon_{s\alpha}(\p) \right]
		\left[p_\lambda \varepsilon_{s\beta}(\p)- p_\beta \varepsilon_{s\lambda}(\p) \right]
		\bar{q}_\gamma \bar{q}_\delta 	q^0 n_\varphi(\x)K^\mu p^0\rho^\gamma(\mathbf{p}) ~. \nonumber\\
	\end{eqnarray}
	We are working in the Coulomb gauge in which the polarization vector of the photon is perpendicular to the direction of propagation ($p\cdot\varepsilon_{s}=0$). Therefore, the above relation reduces to
	\begin{eqnarray}
		\Delta \bar{k}^{(1)\mu}
		&=&i\frac{\kappa^2}{4}\int d\tau d\tau' d\p dK D^{\alpha\beta\gamma\delta}(K)e^{-(\sigma_\gamma^2+\sigma_\varphi^2)|\K|^2/2}e^{-iK\cdot (x(\tau)-x'(\tau'))}
		\nonumber \\ &&\times~V\,V'\,
		\left[-p_\alpha p_\beta \right]
		\bar{q}_\gamma \bar{q}_\delta 	q^0 n_\varphi(\x)K^\mu\rho^\gamma(\mathbf{p}) p^0~. 
	\end{eqnarray}
	This deflection can be considered in the interaction of a light beam with a massive object. As in Section~\ref{sec:massive}, we consider one massive object ($V'n_\varphi(\x)=1$) and the number of photons inside the coherent beam of light is $\mathcal{N}_\gamma=V\,n_\gamma(\x)$. Therefore, using Eq.~\eqref{eq:photonrho} the above equation reduces to
	\begin{eqnarray}
		\Delta \bar{k}^{(1)\mu}
		&=&-8\pi G\int \frac{d^4K}{(2\pi)^4}iK^\mu \bar{p}_\alpha \bar{p}_\beta \bar{q}_\gamma \bar{q}_\delta D^{\alpha\beta\gamma\delta}(K)	 \mathcal{N}_\gamma e^{-(\sigma_\gamma^2+\sigma_\varphi^2)|\K|^2/2}\nonumber\\
		&&~\times\int d\sigma \int d\sigma'e^{-iK\cdot (x(\sigma)-x'(\sigma'))}~, 
	\end{eqnarray}
	where we have used $\sigma=\tau/k^0$ and $\sigma'=\tau'/q^0$. The photon wavelength, $\lambda$, must be much smaller than the distance between the impact parameter $b$
	\begin{eqnarray}\label{eq:lambdab}
		\lambda\ll b~.
	\end{eqnarray}
	This is the regime where geometric optics applies, which is also relevant to Eddington's observations of light deflection. Having a collimated light beam requires
	\begin{eqnarray}
		\lambda\ll w_\gamma~,
	\end{eqnarray}
	where $w_\gamma$ is the beam waist radius that should be smaller than the impact parameter ($w_\gamma\ll b$). On the other hand, the photon wavelength cannot be too small either. The wavelength needs to be much larger than the Compton wavelength, $\lambda_C$, of the massive object
	\begin{eqnarray}\label{eq:lambdacompt}
		\lambda\gg\lambda_C~.
	\end{eqnarray}
	The above relations (Eqs.~\eqref{eq:lambdab}-\eqref{eq:lambdacompt}) lead to the following integrated condition which is similar to the Goldilock's condition in the previous section
	\begin{eqnarray}
		\lambda_C\ll\lambda\ll w_\gamma\ll b~.
	\end{eqnarray}
	This careful balance between size scales ensures our light beam behaves precisely as required for calculating the deflection due to gravity.

	\section{Conclusion and discussion}
	\label{sec:conc}
	
	Our work presented a new method for calculating gravitational impulse exerted by a massive object on another massive object or a light beam. The method is based on QBE, a robust framework for studying the operator evolution of a system interacting with an environment. We analyzed the gravitational impulse as follows. We described the massive objects $1$ and $2$ using two scalar fields. The object $1$ as a system interacting gravitationally with object $2$ as the environment. In the case of the impulse on a photon beam, we considered the photon as a system and a massive object as the environment. The time evolution of the momentum difference of the system or impulse can be derived from the forward scattering term of QBE.
	
	Notably, we have found that when the initial states are represented as plane waves, the forward scattering term and thus the first order gravitational impulse vanishes. In contrast, adopting Gaussian wave packets as initial states yields a non-vanishing impulse in the forward scattering term. The second order problem contributed to the collision term of the QBE have been also calculated with the wavepacket approach and the result of both forward scattering and collision term are consistent with previous results obtained by different approaches. Moreover, the use of the Gaussian wave packet in our new approach has another interesting aspect, which is that the Goldilocks condition is entirely determined. This condition bounds the Gaussian wave packet length scale of two scalar fields between Compton wavelength and scattering length scale. A similar Goldilocks condition is obtained in the case of a light beam interacting with a massive object. In this case, the wavelength of the photon should be limited between the Compton wavelength of the massive object and the impact parameter of the interaction.
	
	This formulation enhances the representations offered by the scattering amplitude and effective field theory (EFT) approaches, which are now widely used tools for studying gravitational interactions \cite{Cheung:2018wkq,Bjerrum-Bohr:2018xdl}. The forward scattering result that obtained here for Gaussian wavepackets, provides a new understanding within this landscape of theoretical tools. Amplitude-based and effective field theory (EFT) approaches have achieved high Post Minkowskian (PM) accuracy in modeling conservative and radiative gravitational dynamics~\cite{Bern:2019nnu,Kalin:2020mvi,Bern:2021ppb}. To reach that level of accuracy and demonstrate the power of our approach, future works can extend the calculation within the forward scattering theorem to include higher order loop contributions in the interaction Hamiltonian. This involves considering processes where particles interact via graviton exchange up to the one loop stage and potentially higher loops. Techniques for multiloop integration developed in collider physics are now being applied to gravitational scattering problems ensuring the computational feasibility of them. Beyond the forward scattering analysis, exploring the collision term in future work would lead to move beyond the current level of approximation and calculate higher refinements to the momentum impulse and scattering (eikonal) angle. Furthermore, using the field theory techniques employed in calculating the interaction Hamiltonian, our approach can be generalized from scalar particles to particles with spin. Various formalisms, including the use of arbitrary-spin field theories and massive spinor helicity, have been developed to incorporate spin effects like spin-orbit and spin-spin interactions \cite{Khriplovich:1997ni,Bern:2020buy,FebresCordero:2022jts,Guevara:2018wpp,Liu:2021zxr,Maia:2017yok,Maia:2017gxn,Bern:2023ity,Belenchia2018,Danielson2022gravitationally}. Applying our forward scattering calculation to such particles would lead to derive spin-dependent scattering results, such as the momentum impulse and spin kick.
	
	The applications of such calculations are directly relevant to modern gravitational wave astronomy. Precise knowledge of conservative and radiative dynamics, including spin effects at high post Newotonian and PM orders, is essential for modeling binary inspirals of compact objects like black holes and neutron stars. Accurate templates are necessary not only for detection but also for precise parameter estimation of astrophyiscal sources. Extending our method to higher orders and spinning particles can contribute valuable new results to this effort, potentially offering an independent check on calculations from other formalisms.

	\section*{Acknowledgments}
	
	We would like to thank M. Abdi for useful discussions and comments. MZ would like to thank INFN and department of Physics and Astronomy “G. Galilei” at University of Padova and also the CERN Theory Division for warm hospitality while this work was done. NB and SM acknowledge financial support from the INFN InDark initiative. This work is supported in part by the MUR Departments of Excellence grant ``Quantum Frontiers''.
	
	\appendix
	\section{Equivalency of KMOC and Damour models for impulse}\label{appendix:KMOCDAMOUR}
	Kosower et al.~\cite{Kosower:2018adc} have obtained the leading-order impulse given in Eq.~\eqref{eq:KMOC}. Its leading order scattering amplitude is
	\begin{eqnarray}\label{eq:A}
		\bar{\mathcal{A}}^{(0)}\left(p, q \rightarrow p+\hbar \bar{K}, q-\hbar \bar{q}\right)=p_{\mu} p_{\nu} D^{\alpha\beta\alpha'\beta'}(K) q_{\alpha} q_{\beta}~.
	\end{eqnarray}
	Putting \eqref{eq:A} into \eqref{eq:KMOC} leads to
	\begin{eqnarray}\label{eq:KMOC2}
		\Delta p^{\mu,(0)} = \frac{2 G}{\pi} \int d^4 \bar{K}\,p_{\alpha'} p_{\beta'} D^{\alpha'\beta'\alpha\beta}(K) q_{\alpha}q_{\beta} \delta\left(\bar{K} \cdot p\right) \delta\left(\bar{K} \cdot q\right) e^{-i b \cdot \bar{K}} i\bar{K}^\mu  .
	\end{eqnarray}
	Furthermore, Damour~\cite{Damour:2016gwp} has derived \eqref{Dp2} and the following relation in his computation holds
	\begin{eqnarray}
		\int d \sigma_1 \int d \sigma_2  \, e^{ i K\cdot (x_1(\sigma_1) - x_2(\sigma_2))}=(2\pi)^2\delta\left(K \cdot p\right) \delta\left(K \cdot q\right)e^{-i K\cdot b}~.
	\end{eqnarray}
	Using the above relation changes Eq.~\eqref{Dp2} into
	\begin{eqnarray} 
		\Delta p_{ \mu}&=& 8 \pi G \int  \frac{d^4 K}{(2\pi)^4} i K_{\mu} p_{\alpha'} p_{\beta'} D^{\alpha'\beta'\alpha\beta}(K) q_{\alpha}q_{\beta}
		(2\pi)^2\delta\left(K\cdot p\right) \delta\left(K \cdot q\right)e^{-i K\cdot b}\nonumber\\
		&=& \frac{2 G}{\pi} \int  d^4 K \,i K_{\mu} p^{\alpha} p_{\alpha'} p_{\beta'} D^{\alpha'\beta'\alpha\beta}(K) q_{\alpha}q_{\beta}
		\delta\left(K \cdot p\right) \delta\left(K \cdot q\right)e^{-i K\cdot b }\,,
	\end{eqnarray}
	that is the same as Eq.~\eqref{eq:KMOC2} by Kosower et al~\cite{Kosower:2018adc} for the leading order scattering amplitude.
	
	\section{Current and energy-momentum tensors} \label{appendix:currentandstress}
	
	Let us first show explicitly how a finite width of the current world tubes can be implemented in our formalism. In the next section, we will demonstrate how this implementation gives rise to an ultraviolet cutoff scale.
	
	For a relativistic massive particle, we can define the stress-energy tensor $T^{\mu\nu}$ and the current $j^\mu$ as follows.
	Taking into account Eq.~\eqref{eq:delta4} one can write 
	\begin{eqnarray}
		T^{\mu\nu}=m\int_{-\infty}^{\infty}d\tau u^\mu u^\nu \delta^{4}(x-x(\tau))=\int_{-\infty}^{\infty}d\tau \frac{p^\mu p^\nu}{m} \delta^{4}(x-x(\tau))~.
	\end{eqnarray}
	
	The integral is taken along the particle's worldline.
	This expression essentially integrates the energy density and momentum of the particle along its worldline, weighted by the particle's four-velocity. 
	
	We begin with the current density of particle 1 which is taken to be concentrated within world tubes of spatial extent $\sigma_\chi$ around the
	world lines $\bar{x}$
	\beq \label{jmu1}
	j^\mu(x)=\int d\tau \mean{\hat{j}^\mu(x) }V\delta^4_{\sigma_\chi}(x-\bar{x}(\tau))~,
	\eeq
	where $\delta^4_{\sigma_\chi}(x-\bar{x}(\tau))$ is defined in Eq.~\eqref{eq:delta4bar} and $\hat{j}^\mu(x)$ is an operator given by
	\beq
	\hat{j}^\mu(x)=2i\hat{\chi}^-(x)\partial^\mu \hat{\chi}^+(x)~.
	\eeq
	The expectation value in \eqref{jmu1} is defined as Eq.~\eqref{eq:expected}. Now, using \eqref{chi1} and \eqref{eq:expda} we get
	\begin{eqnarray}
		j^\mu(x)&=&\int d\tau d\p d\p' p^\mu  \mean{\hat{a}^\dag(\p')\hat{a}(\p)}e^{-ip\cdot x}e^{ip'\cdot x}V\delta^4_{\sigma_\chi}(x-\bar{x}(\tau))
		\nonumber \\ &=&
		\int d\tau d\p d\p' p^\mu  (2\pi)^3 p^0 \rho^\chi(\p)\delta^3(\p-\p') e^{-ip\cdot x}e^{ip'\cdot x}V\delta^4_{\sigma_\chi}(x-\bar{x}(\tau))
		\nonumber \\ &=&
		\int d\tau d\p  p^\mu  \rho^\chi(\p) V\delta^4_{\sigma_\chi}(x-\bar{x}(\tau))
		\nonumber \\ &=&
		\int d \tau \frac{\bar{p}^\mu}{\bar{p}^0} \,n_\chi(\x)V\delta^4_{\sigma_\chi}(x-\bar{x}(\tau))
		=
		\int d \sigma  \bar{p}^\mu\, n_\chi(\x) V\delta^4_{\sigma_\chi}(x-\bar{x}(\sigma))
		~,
	\end{eqnarray}
	where we have used
	\beq
	\int  d\p  p^\mu  \rho^\chi(\p) =\frac{\bar{p}^\mu} {p^0}n_\chi(\x)~,
	\eeq
	and $\sigma=\tau/p^0$. 
	
	The stress energy momentum tensor for particle 1 is also given by 
	\begin{eqnarray}
		T^{\mu\nu}(x)&=&
		\int d\tau d\p d\p' p^\mu p^\nu \mean{\hat{a}^\dag(\p')\hat{a}(\p)}e^{-ip\cdot x}e^{ip'\cdot x}V\delta^4_{\sigma_\chi}(x-\bar{x}(\sigma))
		\nonumber \\ &=&
		\int d\sigma\, \bar{p}^\mu \bar{p}^\nu n_\chi(\x)V\delta^4_{\sigma_\chi}(x-\bar{x}(\sigma))~.
	\end{eqnarray}
	which is manifestly symmetric and Lorentz covariant.

	\section{Derivation of the second-order collision integral}
	\label{C}
	
	In this appendix, we present the detailed calculation of the second-order ($\mathcal(O)(G^2)$
	collision integral for the gravitational impulse 
	$\Delta \bar{k}^{(2)\mu}$, as introduced in Eq. \eqref{QBEkmu-1-1} of the main text. Starting from the double-commutator structure of the collision term of the QBE, we systematically calculate the expectation values of scalar field operators, perform phase-space integrations, and then resolve momentum constraints. 
	We begin by taking the integration over the spacetime coordinates 
	$x_1$, $x'_1$, $x_2$ and $x'_2$ which yields
	\begin{eqnarray} 
		\Delta \bar{k}^{(2)\mu} &=&
		-\frac{\kappa^4}{32}V^2\,V'^2\int 
		d\tau_1 d\tau'_1 d\tau_2 d\tau'_2
		d\p_1 d\p'_1  d\q_1 d\q'_1	d\p_2 d\p'_2  d\q_2 d\q'_2
		dK_1 dK_2  p_{1\alpha_1} p'_{1\beta_1} q_{1\alpha'_1}q'_{1\beta'_1}  
		p_{2\alpha_2} p'_{2\beta_2} 
		\nonumber \\  &   \times & 
		q_{2\alpha'_2}q'_{2\beta'_2}
		D^{\alpha_1\beta_1\alpha'_1\beta'_1}(K_1) D^{\alpha_2\beta_2\alpha'_2\beta'_2}(K_2) 
		e^{-i(K_1+p_1-p'_1)\cdot x_1(\tau_1)}e^{-i(q_1-q'_1-K_1)\cdot x_1(\tau'_1)} 	e^{-i(K_2+p_2-p'_2)\cdot x_2(\tau_2)}
		\nonumber \\  &   \times & 
		e^{-i(q_2-q'_2-K_2)\cdot x_1(\tau'_2)}
		e^{-\sigma_\chi^2|\K_1+\p_1-\p'_1|^2/2}e^{-\sigma_\varphi^2|\K_1+\q'_1-\q_1|^2/2}e^{-\sigma_\chi^2|\K_2+\p_2-\p'_2|^2/2}
		e^{-\sigma_\varphi^2|\K_2+\q'_2-\q_2|^2/2}
		\nonumber \\  &   \times & 
		\Big [
		\left<\hat{b}^\dag(\q'_1)\hat{b}(\q_1)\hat{b}^\dag(\q'_2)\hat{b}(\q_2) \right >  \Big [	\left<\hat{a}^\dag(\p'_1)\hat{a}(\p_1)\hat{a}^\dag(\p'_2)\hat{a}(\p_2) k^\mu \right > -	\left< \hat{a}^\dag(\p'_1)\hat{a}(\p_1)k^\mu\hat{a}^\dag(\p'_2)\hat{a}(\p_2) \right > \Big ]
		\nonumber \\  &  - & 
		\left<\hat{b}^\dag(\q'_2)\hat{b}(\q_2)\hat{b}^\dag(\q'_1)\hat{b}(\q_1) \right >  \Big [	\left<\hat{a}^\dag(\p'_2)\hat{a}(\p_2) k^\mu \hat{a}^\dag(\p'_1)\hat{a}(\p_1)\right > -	\left<k^\mu\hat{a}^\dag(\p'_2)\hat{a}(\p_2)\hat{a}^\dag(\p'_1)\hat{a}(\p_1) \right > \Big]  \Big]~.
		\nonumber 
		\\
	\end{eqnarray}
	Next, we insert \eqref{kmu} in the above expectation values. To evaluate them we use the following expresions \cite{Zarei:2021dpb}
	\begin{eqnarray} 
		\left<\hat{a}^\dag(\p'_1)\hat{a}(\p_1)\hat{a}^\dag(\p'_2)\hat{a}(\p_2) \hat{a}^\dag(\k)\hat{a}(\k) \right >  &\sim& \delta^3(\p'_1-\k)\delta^3(\p'_2-\p_2)\rho^\chi(\k)\rho^\chi(\p_2)\delta^3(\k-\p_1)[1+\rho^\chi(\p_1)]
		\nonumber \\ 
		&+&
		\delta^3(\p'_1-\p_2)\delta^3(\p'_2-\k)\rho^\chi(\k)\rho^\chi(\p_2)\delta^3(\k-\p_1)[1+\rho^\chi(\p_1)]
		\nonumber \\ 
		&+&
		\delta^3(\p'_1-\p_2)\delta^3(\k-\k)\rho^\chi(\k)\rho^\chi(\p_2)\delta^3(\p'_2-\p_1)[1+\rho^\chi(\p_1)]
		\nonumber \\ 
		&+&
		\delta^3(\p'_1-\p_1)\delta^3(\p'_2-\k)\rho_\chi(\k)\rho^\chi(\p_1) \delta^3(\k-\p_2)[1+\rho^\chi(\p_2)]
		\nonumber \\ 
		&+&
		\delta^3(\p'_1-\p_1)\delta^3(\k-\k)\delta^3(\p'_2-\p_2)\rho^\chi(\k)\rho^\chi(\p_2)\rho^\chi(\p_1) 
		~,
		\nonumber \\
	\end{eqnarray}
	
	\begin{eqnarray} 
		\left< \hat{a}^\dag(\p'_1)\hat{a}(\p_1)\hat{a}^\dag(\k)\hat{a}(\k)\hat{a}^\dag(\p'_2)\hat{a}(\p_2) \right >
		&\sim& \delta^3(\p'_1-\p_2)\delta^3(\k-\k)\rho^\chi(\k)\rho^\chi(\p_2)\delta^3(\p'_2-\p_1)[1+\rho^\chi(\p_1)]
		\nonumber \\ 
		&+&
		\delta^3(\p'_1-\k)\delta^3(\p'_2-\p_2)\rho^\chi(\k)\rho^\chi(\p_2)\delta^3(\k-\p_1)[1+\rho^\chi(\p_1)]
		\nonumber \\ 
		&+&
		\delta^3(\p'_1-\k)\delta^3(\k-\p_2)\rho^\chi(\k)\rho^\chi(\p_2)\delta^3(\p'_2-\p_1)[1+\rho^\chi(\p_1)]
		\nonumber \\ 
		&+&
		\delta^3(\p'_1-\p_1)\delta^3(\k-\p_2)\rho^\chi(\p_1)\rho^\chi(\p_2)\delta^3(\p'_2-\k)[1+\rho^\chi(\k)]
		\nonumber \\ 
		&+&
		\delta^3(\p'_1-\p_1)\delta^3(\k-\k)\delta^3(\p'_2-\p_2)\rho^\chi(\k)\rho^\chi(\p_2)\rho^\chi(\p_1) ~,
		\nonumber \\ 
	\end{eqnarray}
	
	\begin{eqnarray} 
		\left< \hat{a}^\dag(\p'_2)\hat{a}(\p_2)\hat{a}^\dag(\k)\hat{a}(\k)\hat{a}^\dag(\p'_1)\hat{a}(\p_1) \right >
		&\sim&  \delta^3(\p'_2-\p_1)\delta^3(\k-\k)\rho^\chi(\k)\rho^\chi(\p_1)\delta^3(\p'_1-\p_2)[1+\rho^\chi(\p_2)]
		\nonumber \\ 
		&+&
		\delta^3(\p'_2-\k)\rho^\chi(\k)\delta^3(\p_1-\k)\rho^\chi(\p_1)\delta^3(\p'_1-\p_2)[1+\rho^\chi(\p_2)]
		\nonumber \\ 
		&+&
		\delta^3(\p'_2-\k)\rho^\chi(\k)\delta^3(\p_1-\p'_1)\rho^\chi(\p_1)\delta^3(\k-\p_2)[1+\rho^\chi(\p_2)]
		\nonumber \\ 
		&+&
		\delta^3(\p'_2-\p_2)\rho^\chi(\p_2)\delta^3(\p_1-\k)\rho^\chi(\p_1)\delta^3(\p'_1-\k)[1+\rho^\chi(\k)]
		\nonumber \\ 
		&+&
		\delta^3(\p'_2-\p_2)\delta^3(\k-\k)\delta^3(\p'_1-\p_1)\rho^\chi(\k)\rho^\chi(\p_1)\rho^\chi(\p_2) ~,
		\nonumber \\ 
	\end{eqnarray}
	and
	\begin{eqnarray} 
		\left<\hat{a}^\dag(\k)\hat{a}(\k)\hat{a}^\dag(\p'_2)\hat{a}(\p_2)\hat{a}^\dag(\p'_1)\hat{a}(\p_1) \right >
		&\sim&  \delta^3(\k-\p_1)\delta^3(\p'_2-\p_2)\rho^\chi(\p_1)\rho^\chi(\p_2)\delta^3(\p'_1-\k)[1+\rho^\chi(\k)]
		\nonumber \\ 
		&+&
		\delta^3(\k-\p_2)\rho^\chi(\p_2)\delta^3(\p'_2-\p_1)\rho^\chi(\p_1)\delta^3(\p'_1-\k)[1+\rho^\chi(\k)]
		\nonumber \\ 
		&+&
		\delta^3(\k-\p_2)\delta^3(\p'_1-\p_1)\rho^\chi(\p_1)\rho^\chi(\p_2)\delta^3(\p'_2-\k)[1+\rho^\chi(\k)]
		\nonumber \\ 
		&+&
		\delta^3(\k-\k)\rho^\chi(\k)\delta^3(\p'_2-\p_1)\rho^\chi(\p_1) \delta^3(\p'_1-\p_2)[1+\rho^\chi(\p_2)]
		\nonumber \\ 
		&+&
		\delta^3(\k-\k)\delta^3(\p'_2-\p_2)\delta^3(\p'_1-\p_1)\rho^\chi(\k)\rho^\chi(\p_2)\rho^\chi(\p_1) ~.
		\nonumber \\ 
	\end{eqnarray}
	Note that we have
	omitted kinematic factors (e.g., $(2\pi)^3$, $p^0$, ... ) in these expresions. These neglected factors are restored during the following calculations. The same expresion is given for expectation values of the operators $\hat{b}(\q)$ \cite{Zarei:2021dpb}.  
	Now, using the Dirac delta functions appeared in expectation values we integrate over 
	$\q'_1$, $\q'_2$,  $\p'_1$ and $\p'_2$ and find
	We therefore find
	\begin{eqnarray}
		\Delta \bar{k}^{(2)\mu} &=&
		-\frac{\kappa^4}{16}V^2\,V'^2\int 
		d\tau_1 d\tau'_1 d\tau_2 d\tau'_2
		d\p_1   d\q_1 	d\p_2  d\q_2 
		dK_1 dK_2  p_{1\alpha_1} p_{2\beta_1} q_{1\alpha'_1}q_{2\beta'_1}  
		p_{2\alpha_2} p_{1\beta_2} q_{2\alpha'_2}q_{1\beta'_2}
		\nonumber \\  &   \times & 
		D^{\alpha_1\beta_1\alpha'_1\beta'_1}(K_1) D^{\alpha_2\beta_2\alpha'_2\beta'_2}(K_2) 
		e^{-i(K_1+p_1-p_2)\cdot x_1(\tau_1)}e^{-i(q_1-q_2-K_1)\cdot x_1(\tau'_1)} 	e^{-i(K_2+p_2-p_1)\cdot x_2(\tau_2)}
		\nonumber \\  &   \times & 
		e^{-i(q_2-q_1-K_2)\cdot x_2(\tau'_2)}
		e^{-\sigma_\chi^2|\K_1+\p_1-\p_2|^2/2}e^{-\sigma_\varphi^2|\K_1+\q_2-\q_1|^2/2}e^{-\sigma_\chi^2|\K_2+\p_2-\p_1|^2/2}
		e^{-\sigma_\varphi^2|\K_2+\q_1-\q_2|^2/2}
		\nonumber \\  &   \times & 
		\rho^\varphi(\q_1) \rho^\varphi(\q_2)\Big [ 	p^\mu_1 \rho^\chi(\p_1)(\rho^\chi(\p_1)+\rho^\chi(\p_2)
		)	-
		p^\mu_2 \rho^\chi(\p_2)(\rho^\chi(\p_1) 
		+\rho^\chi(\p_2))
		\Big ]~.
	\end{eqnarray}
	Finally, we integrate over $\q_1$, $\q_2$, $\p_1$, and $\p_2$. This leads to two Gaussian expressions, $\delta_{\sigma_\varphi}(\tau_1-\tau_2)$ and $\delta_{\sigma_\varphi}(\tau'_1-\tau'_2)$, which can then be used to eliminate the variables $\tau_1$ and $\tau'_1$. Consequently, we find
	\begin{eqnarray}
		\Delta \bar{k}^{(2)\mu} &\approx&
		\frac{\kappa^4}{16}\int 
		d\sigma_2 d\sigma'_2
		dK_1 dK_2
		e^{-\sigma_\chi^2(|\K_1|^2+|\K_2|^2)/2}e^{-\sigma_\varphi^2(|\K_1|^2+|\K_2|^2)/2} 
		e^{-i(K_1+K_2)\cdot ( x_2(\sigma_2)-x_2(\sigma'_2))}
		\nonumber \\  &   \times & 	(K_1^\mu-K_2^\mu) \,
		[ \bar{p}_{1\alpha_1} \bar{p}_{2\beta_1}
		D^{\alpha_1\beta_1\alpha'_1\beta'_1}(K_1)
		\bar{q}_{1\alpha'_1}\bar{q}_{2\beta'_1} ]\, [
		\bar{p}_{2\alpha_2} \bar{p}_{1\beta_2} 
		D^{\alpha_2\beta_2\alpha'_2\beta'_2}(K_2) 
		\bar{q}_{2\alpha'_2}\bar{q}_{1\beta'_2}]\,.
	\end{eqnarray}

	
	\nocite{apsrev41Control}
	\bibliographystyle{apsrev4-1}
	\bibliography{refs.bib,refcontrol.bib}


	
\end{document}